\documentclass[%
 reprint,
 floatfix,
 amsmath,amssymb,
 aps,
]{revtex4-1}

\usepackage{graphicx}% Include figure files
\usepackage{dcolumn}% Align table columns on decimal point
\usepackage{bm}% bold math
\usepackage{color}
\usepackage{lineno}
\usepackage{hyperref}% add hypertext capabilities
\newcommand{\jim}[1]{\textcolor{black}{#1}}

\def\etal{{\it et al.}}

\def\eg{{\it e.g.}}

\def\Fig#1{Fig.~\ref{#1}}

\def\ee{e^+e^-} 

\begin{document}

\preprint{LCCPEB---}

\title{The International Linear Collider \\ A Global Project}% Force line breaks with \\
%\thanks{Version 4.0}%

\author{\textbf{Prepared by:}
Hiroaki Aihara$^1$, Jonathan Bagger$^2$, Philip Bambade$^3$, Barry Barish$^4$,  Ties Behnke$^5$, Alain Bellerive$^6$, Mikael Berggren$^5$, James Brau$^7$, Martin Breidenbach$^8$, 
Ivanka Bozovic-Jelisavcic$^9$,
Philip Burrows$^{10}$, Massimo Caccia$^{11}$, Paul Colas$^{12}$, Dmitri Denisov$^{13}$, Gerald Eigen$^{14}$, Lyn Evans$^{15}$, Angeles Faus-Golfe$^{3}$, Brian Foster$^{5,10}$, Keisuke Fujii$^{16}$, Juan Fuster$^{17}$, Frank Gaede$^{5}$, Jie Gao$^{18}$, Paul Grannis$^{19}$, Christophe Grojean$^{5}$, Andrew Hutton$^{20}$, Marek Idzik$^{21}$, Andrea Jeremie$^{22}$, Kiyotomo Kawagoe$^{23}$, Sachio Komamiya$^{1,24}$, Tadeusz Lesiak$^{25}$, Aharon Levy$^{26}$, Benno List$^{5}$, Jenny List$^{5}$, Shinichiro Michizono$^{16}$, Akiya Miyamoto$^{16}$, Joachim Mnich$^{5}$, Hugh Montgomery$^{20}$, Hitoshi Murayama$^{27}$, Olivier Napoly$^{12}$, Yasuhiro Okada$^{16}$, Carlo Pagani$^{28}$, Michael Peskin$^{8}$, Roman Poeschl$^{3}$, Francois Richard$^{3}$, Aidan Robson$^{29}$, Thomas Schoerner-Sadenius$^{5}$, Marcel Stanitzki$^5$, Steinar Stapnes$^{15}$, Jan Strube$^{7,30}$, Atsuto Suzuki$^{31}$, Junping Tian$^{1}$, Maksym Titov$^{12}$, Marcel Vos$^{17}$, Nicholas Walker$^{5}$, Hans Weise$^{5}$, Andrew White$^{32}$, Graham Wilson$^{33}$, Marc Winter$^{34}$, Sakue Yamada$^{1,16}$, Akira Yamamoto$^{16}$, Hitoshi Yamamoto$^{35}$ and Satoru Yamashita$^{1}$. }
% \altaffiliation[Also at ]{Physics Department, XYZ University.}%Lines break automatically or can be forced with \\
%\author{et.al.}%
% \email{Second.Author@institution.edu}
\affiliation{\vspace{.2 cm}$^1$U. Tokyo, $^2$TRIUMF, $^3$LAL-Orsay/CNRS, $^4$Caltech, $^5$DESY, $^6$Carleton U., $^7$U. Oregon, $^8$SLAC, $^9$INN VINCA, Belgrade, $^{10}$Oxford U., $^{11}$U. Insubria, $^{12}$CEA/Irfu, U. Paris-Saclay, $^{13}$Fermilab, $^{14}$U. Bergen, $^{15}$CERN, $^{16}$KEK, $^{17}$IFIC, U. Valencia-CSIC, $^{18}$IHEP, $^{19}$Stony Brook U., $^{20}$Jefferson Lab, $^{21}$AGH, Krak\'ow, $^{22}$LAPP/CNRS, $^{23}$Kyushu U., $^{24}$Waseda U., $^{25}$IFJPAN, Krak\'ow, $^{26}$Tel Aviv U., $^{27}$U. California, Berkeley, $^{28}$INFN, $^{29}$U. Glasgow, $^{30}$PNNL, $^{31}$Iwate Prefecture U., $^{32}$U. Texas, Arlington, $^{33}$U. Kansas, $^{34}$IPHC/CNRS, $^{35}$U. Tohoku }

 % Authors' institution and/or address\\
% This line break forced with \textbackslash\textbackslash
%}%
\collaboration{\textbf{\textit{Representing the Linear Collider Collaboration and the global ILC community.}}}%\noaffiliation

\date{
%$^1$U. Tokyo, $^2$Triumf \\[2ex]
   \today}% It is always \today, today,
             %  but any date may be explicitly specified

\begin{abstract}
%Input from the International Linear Collider community for the European Strategy Update 
\centerline{\textbf{Abstract}}
A large, world-wide community of physicists is working to realise
an exceptional physics program
of energy-frontier, electron-positron collisions with the International Linear Collider
(ILC).  This program will begin with a
central focus on high-precision and model-independent measurements of the Higgs boson couplings.   This method of searching for 
new physics beyond the Standard Model is orthogonal to and complements 
the LHC physics program.   The ILC
at 250~GeV will also search for direct new physics in exotic Higgs decays
and in pair-production of weakly interacting particles.
Polarised electron
and positron beams add unique opportunities to the physics reach.
The ILC can
be upgraded to higher energy, enabling precision studies of the top quark and
measurement of the top Yukawa coupling and the Higgs self-coupling.

%The ILC accelerator is based on the technology of superconducting radio-frequency
%cavities.   This technology has reached a mature stage in 
%the European XFEL and LCLS-II projects.  
%Following decades of technical development, R\&D,
%and design optimisation, the project is ready for construction.  
The key accelerator technology, superconducting radio-frequency cavities,
has matured.
Optimised collider and detector designs, and associated 
physics analyses, were presented in the ILC  Technical Design Report,
signed by 2400 scientists.  

There is a strong interest in Japan to host this
international effort.  A detailed review of the many aspects of the
project is nearing a conclusion in Japan.  Now the Japanese government is preparing for
a decision on the next phase of international negotiations, that could lead to a 
project start within a few years. The potential timeline of the
 ILC project includes an initial phase of about 4 years to obtain
 international agreements, complete engineering design and prepare construction, and form 
 the requisite international
collaboration, followed by a construction phase of 9 years.

\end{abstract}

\pacs{Valid PACS appear here}% PACS, the Physics and Astronomy
                             % Classification Scheme.
%\keywords{Suggested keywords}%Use showkeys class option if keyword
%display desired
\maketitle
\onecolumngrid
\textbf{Supporting documents web page:} 

\vspace{-0.8cm}
%https://linearcollider.web.cern.ch/content/ilc-european-strategy-document 
%\vspace{-1.0cm}
\url{https://ilchome.web.cern.ch/content/ilc-european-strategy-document}

\textbf{Contact persons:} 
James Brau (jimbrau@uoregon.edu), Juan Fuster(Juan.Fuster@ific.uv.es), Steinar Stapnes (Steinar.Stapnes@cern.ch)
\vspace{1cm}

\pagebreak

\pagestyle{plain}
\setcounter{page}{1}

\twocolumngrid
%https://linearcollider.web.cern.ch/content/ilc-european-strategy-document

%\tableofcontents

\vspace{-.4cm}

\section{\label{sec:intro}Introduction}

\vspace{-.3cm}

%\todo{ 1 page - Jim and Juan
% Introduce the ILC250, brief overview of status (technical maturity and TDR, staging, cost analysis, status of political situation) }

A central issue in particle physics today is the search for new
 phenomena needed to address shortcomings of the 
highly successful Standard Model (SM).  These new effects can manifest
themselves as new particles, new forces, 
 or deviations in the predictions of the SM derived from
 high-precision measurements. While the SM is theoretically
 self-consistent,
it leaves many issues of particle physics unaddressed. 
It has no place for the dark matter and dark energy observed in the
cosmos,
and it cannot explain the excess of matter over antimatter.   It has
nothing to say about the mass scale of quarks, leptons, and Higgs and gauge 
bosons, which is much less than the Planck scale.   It has nothing 
to say about the large mass ratios among these particles.   These and 
other issues motivate intense efforts to challenge the predictions of
the SM and search for clues to what lies beyond it.
 
The Higgs boson,
 discovered in 2012 at the Large Hadron Collider (LHC), is
central to the SM,  since it is the  origin of 
 electroweak symmetry-breaking and gives mass to all 
known elementary particles.   The study of the properties and
interactions of the Higgs boson is thus of utmost importance.

The International Linear Collider (ILC) has the capabilities needed
to address these central physics issues.  It will extend and complement the LHC physics program. First and most importantly, it provides
unprecedented precision in the measurements and searches needed to detect deviations from the SM.
Already in its first stage, the 
ILC will have a new level of sensitivity to test the well-defined SM
expectations for the Higgs boson properties, and to advance many other 
tests of SM expectations.  
The well-defined collision energy at the ILC, together with highly polarised beams, low background levels and absence of spectator particles, will enable these precision measurements.
    A linear collider allows straightforward energy
upgrades, which bring new processes into play. 
The energy upgrades will allow the ILC to remain a powerful 
discovery vehicle for decades. 
Finally, and critically, the technology is mature, 
ready for implementation today.

For more than twenty years the worldwide community has been engaged in
a research program to develop the technology required to realise a
high-energy linear collider.  As the linear collider technology has
progressed,
committees of   the International Committee for Future Accelerators
(ICFA) have  guided its successive stages.
In the mid-1990's, as various technology options to
realise a high-energy linear collider were emerging, the 
Linear Collider Technical Review Committee developed a standardised
way to  compare  these  technologies in terms of parameters such as
power consumption and luminosity. In 2002, ICFA set up a second
review panel which concluded that both warm and cold technologies had
developed to the point where either could be the basis for a linear
collider. In 2004, the  International Technology Review Panel
(ITRP) was charged by ICFA to recommend an option and focus the
worldwide R\&D effort.  This panel chose the  superconducting
radiofrequency technology (SCRF), in a large part due to its
energy efficiency and potential for broader applications. 
The effort to design and
establish the technology for the linear collider culminated in the
publication of the Technical Design Report (TDR) for the International
Linear Collider (ILC) in 2013~\cite{Behnke:2013xla}.

The collider design is thus the result of nearly twenty years of
R\&D. The heart of the ILC, the superconducting cavities, is based on
pioneering work of the TESLA Technology Collaboration. Other aspects of the 
technology
emerged from the R\&D carried out for the JLC/GLC and NLC projects,
which were based on room-temperature accelerating structures. From
2005 to the publication of the TDR~\cite{Behnke:2013xla} in 2013, the
design of the ILC accelerator was conducted under the mandate of ICFA
as a worldwide
international collaboration, the Global Design Effort (GDE). 
Since 2013, ICFA has placed the  international activities for both the ILC and CLIC
projects under a single organisation, 
the Linear Collider Collaboration (LCC).
Today the European XFEL provides an operating 1/10-scale demonstration of the fundamental SCRF technology.

Once the mass of the Higgs boson was known, it was established that the
linear collider could start its ambitious physics program with an initial centre-of-mass energy of 250 GeV, with a reduced cost relative to that in the TDR.  In this ILC250~\cite{Evans:2017rvt}, the final focus and beam dumps would be designed to operate at energies up to 1 TeV.
Advances in the theoretical understanding of the impact of precision
measurements at the 
 ILC250 have justified that this operating point already gives
 substantial 
sensitivity to physics beyond the Standard 
Model~\cite{Barklow:2017suo,Fujii:2017vwa}. 
 The cost estimate for ILC250 was also carefully evaluated;
it is described in Appendix A. It is similar in scale to the
LHC 
project.

In its current
form, the ILC250 is a $250\,{\mathrm{GeV}}$ centre-of-mass energy
(extendable up to $1\,{\mathrm{TeV}}$) linear $e^+e^-$ collider, based
on $1.3\,{\mathrm{GHz}}$  SCRF
cavities. It is designed to achieve a luminosity of $1.35\cdot
10^{34}~{\mathrm{cm}}^{-2}{\mathrm{s}}^{-1}$, and provide an integrated
luminosity of $400\,{\mathrm{fb}}^{-1}$ in the first four years of
running and $2\,{\mathrm{ab}}^{-1}$ in a little over a decade. The electron beam will be polarised to $80\,\%$, and the baseline plan includes an 
undulator-based
positron source which will  deliver
$30\,\%$ positron  polarisation. 
Positron production by a 3 GeV S-band injector
is an alternative being considered.

The experimental community has developed
designs for two complementary detectors, ILD and SiD, 
as described in \cite{Behnke:2013lya}. These detectors are designed to 
optimally address the
ILC physics goals, operating in a push-pull configuration.  
The detector R\&D program leading to these designs
has 
contributed a number of advances in 
detector capabilities with applications well beyond the linear
collider program. 

This report summarises the current status of this effort, describing
the physics reach, the technological maturity of the accelerator,
detector, and software/computing designs,
plus a short discussion on the further steps 
 needed to realise the project.

\vspace{-.4cm}

\section{\label{sec:phys}Physics}

\vspace{-.3cm}

The ILC has the ability to begin with a high-precision study of the Higgs boson couplings.  At 250~GeV, the ILC also
presents many opportunities to discover new particles that go beyond
the capabilities of the LHC.  Finally, the ILC at 250~GeV opens the
door to further exploration of $\ee$ reactions at higher energies. This capability has been clearly demonstrated with detailed simulations of important physics channels including full detector effects. The ILC physics case is reviewed at greater length in the reference document~\cite{ILCforESS}.
%, including examples of the detailed detector simulations.

The Higgs boson is a necessary element of the SM, yet it is to very large extent unknown.    In the SM, the Higgs field
couples to every elementary particle and provides the mass for all
quarks, charged leptons, and heavy vector bosons.   The LHC has discovered
the Higgs particle and confirmed the presence of the couplings responsible for the
masses of the $W$, $Z$, $t$, $b$, and $\tau$~\cite{LHCHiggssummary}. 
 However, mysteries are still
buried here.   The Higgs couplings are not universal, as the gauge
couplings are, and their pattern (which is also the pattern of lepton
and quark masses) is not explained by the SM.  The basic phenomenon that provides
mass for elementary particles---the spontaneous breaking of the gauge
symmetry $SU(2)\times U(1)$---is not explained, and actually cannot be
explained, by the SM.   The Higgs boson could also couple to new
particles and fields that have no SM gauge interactions and are
otherwise completely inaccessible to observation.  Thus, detailed
examination of the Higgs boson properties should be a next major
goal for particle physics experiments.

Within the SM, the couplings of the Higgs boson are specified now that
the parameters of the model, including the Higgs boson mass, are
known.  Expected knowledge improvements of SM parameters in the 2020's will allow these couplings to be predicted to the part-per-mille level~\cite{Lepage:2014fla}.
Models of new physics modify these predictions at the 10\% level or less, 
detectable by precision experiments.   Most importantly, different
classes of models affect the various Higgs couplings differently, so that
systematic measurement of the Higgs couplings can reveal clues to the
nature of the new interactions.   The precision study of the
Higgs boson interactions then provides a new method both to {\it discover}  the
presence of physics beyond the SM and to {\it learn}  about its nature.

%%%%%%%%%%%%%%%%
\begin{figure}
\begin{center}
\includegraphics[width=0.99\hsize]{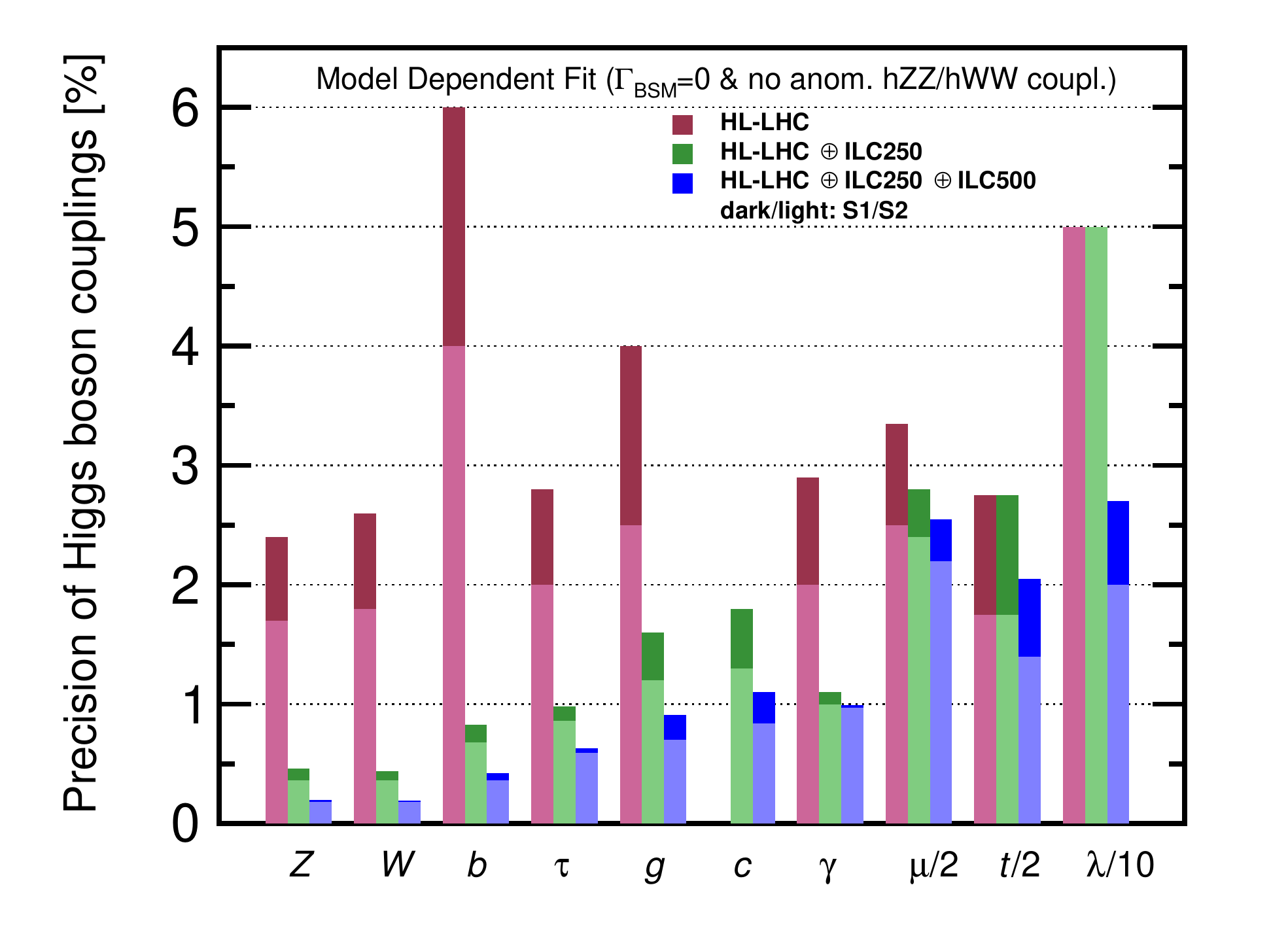}
\caption{Projected Higgs boson coupling uncertainties for the LHC and
  ILC
using the model-dependent assumptions appropriate to the LHC Higgs
coupling fit.   The
dark- and light-red bars represent the projections in the scenarios S1
and S2 presented in \cite{Yellow,CMSYellow}.  The scenario S1 refers to
analyses with our current understanding; the scenario S2 refers to
more optimistic assumptions in which experimental errors decrease with
experience.  The dark- and light-green bars represent the
projections in the ILC scenarios in similar S1 and S2 scenarios defined in \cite{ILCforESS}. 
 The dark- and light-blue bars show the projections for scenarios S1 and S2
when
data from the 500~GeV run of the ILC is included. The same integrated luminosities are assumed as for Figure \ref{fig:ILCmodelindep}. The projected uncertainties in the Higgs couplings to $\mu\mu$, $tt$, and the self-coupling are divided by the indicated factors to fit on the scale of this plot.}
 \label{fig:ILCLHC}
\end{center}
\vspace{-0.7cm}
\end{figure}
%%%%%%%%%%%%%%%%%%%%%%%%%%%%%%%%%%%%%%%%%%%%%%%%%%%%%%%%%%%%%%
%%%%%%%%%%%%%%%%
\begin{figure}
\begin{center}
\includegraphics[width=0.99\hsize]{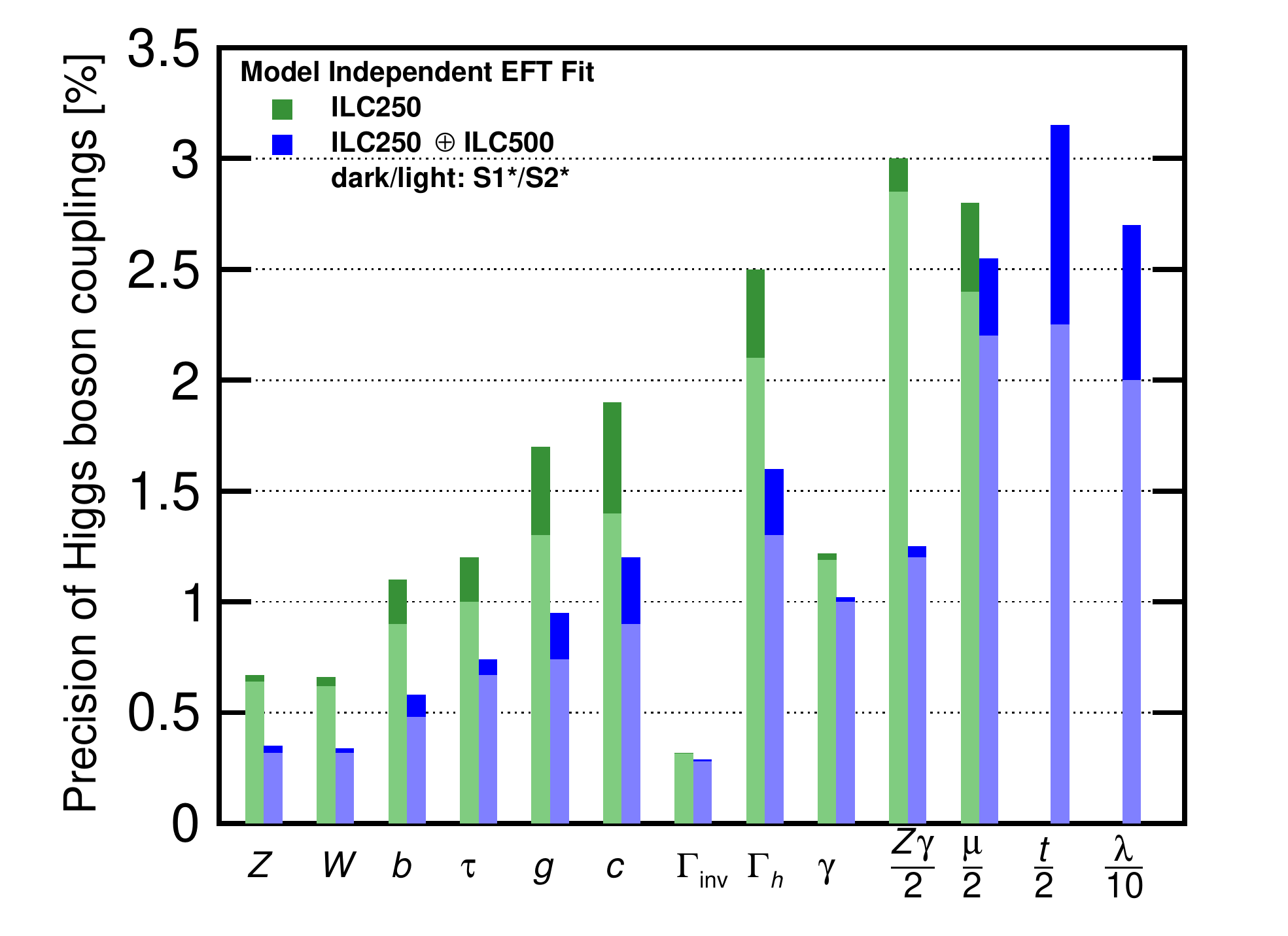}
\caption{Projected Higgs boson coupling uncertainties for the ILC
  program at 250~GeV and an energy upgrade to 500~GeV, using the
  highly model-independent analysis presented in \cite{Barklow:2017suo}. This  analysis makes use of  data on $\ee\to W^+W^-$ in addition to Higgs boson observables and also incorporates projected LHC results, as described  in the text. Results are obtained assuming integrated luminosities of $2\,{\mathrm{ab}}^{-1}$ at 250~GeV and $4\,{\mathrm{ab}}^{-1}$ at 500~GeV. All estimates of uncertainties are derived from full detector simulation. Note that the projected uncertainties in the Higgs couplings to $Z\gamma$, $\mu\mu$, $tt$, and the self-coupling are divided by the indicated factors to fit on the scale of this plot. The scenario S1* refers to analyses with our current understanding; the scenario S2* refers to more optimistic assumptions in which experimental errors decrease with experience.   A full explanation of the analysis and assumptions underlying these estimates is given in \cite{ILCforESS}.}
\label{fig:ILCmodelindep}
\end{center}
\vspace{-0.7cm}
\end{figure}
%%%%%%%%%%%%%%%%%%%%%%%%%%%%%%%%%%%%%%%%%%%%%%%%%%%%%%%%%%%%%%

The couplings of the Higgs boson are now being studied at the LHC.
The LHC experiments have made remarkable progress in measuring the ratios of
couplings of the Higgs boson, and they expect impressive further progress, as
documented in the HL-LHC Yellow Report~\cite{Yellow}.  The uncertainty
projections
from the Yellow Report  are shown in Fig.~\ref{fig:ILCLHC}.   These
measurements are very challenging.   Aside from events in
which the Higgs boson appears as a narrow resonance (the decays to
$\gamma\gamma$ and $4\ell$), Higgs boson events are not visibly
distinct from SM background events. Analyses start from signal/background ratios of about 1/10 even in the most sensitive kinematic regions   (better for VBF production,
but 
worse for $VH$ production with $H \to b\bar b$) and then apply strong
selections to make the Higgs signal visible.   To reach the
performance levels predicted in  the Yellow Report, it is necessary to
determine the level of suppression of SM backgrounds to better than
 1\% accuracy.  At
the same time, these projected uncertainties do not allow the LHC experiments to
observe, for example, an anomaly of 5\% in the $hWW$ coupling to
3$\sigma$ significance.   To prove the presence of such small
deviations, which are typical in new physics models, a
different approach is required. 

What is needed for a precision Higgs boson measurement program is a new experimental method in which all individual Higgs boson decays are manifest and can be studied in detail.   This is provided by the reaction
$\ee\to ZH$ at 250~GeV in the centre-of-mass. At this CM energy, the lab energy spectrum of $Z$ bosons shows a clear peak at 110~GeV, corresponding to recoil against the Higgs boson, on top of a small and precisely calculable SM background. Events in this peak tag the Higgs boson independently of the mode of Higgs boson decay. These events then give a complete picture of Higgs boson  decays, including all SM leptonic and hadronic final states and also invisible or partially visible exotic modes. 

%With small and precisely calculable SM backgrounds, any $Z$ boson
%observed with a lab energy of $\sim$110~GeV is recoiling against a Higgs
%boson.  
  
Further, since the cross section for Higgs production can be measured
independently of any property of the Higgs boson, the total Higgs width and hence the scale of Higgs
couplings can be determined and the individual couplings can be
absolutely normalised.  Each individual coupling can be compared to
its SM prediction.

In the description of new physics by an $SU(2)\times U(1)$-invariant effective field theory (EFT), there exist both a remarkable complementarity and a synergy between measurements in Higgs physics, in precision electroweak observables and in diboson production. This calls for a global approach in interpreting data from the three different sectors. The high precision in  the measurement of $e^+e^- \to W^+W^-$ at an $e^+e^-$ collider then works to improve the Higgs-coupling determination. Beam polarisation at the ILC is also a powerful tool
to separate the contributions of different EFT
coefficients.  In addition, a number of readily interpreted Higgs boson observables that will be measured at the HL-LHC can be used, especially the 
ratio of branching ratios  $BR(H\to \gamma\gamma)/BR(H\to ZZ^*)$. 
  In \cite{Barklow:2017suo}, it is shown that, by the use of this information, 
  it is possible to fit {\it all}
relevant EFT
coefficients  {\it simultaneously}, giving a 
determination of Higgs boson couplings that is as
model-independent as the underlying EFT description itself.

The uncertainties in  cross section and $\sigma\cdot BR$ measurements that contribute to the EFT determination of the Higgs boson couplings were estimated using full-simulation analyses.  These analyses incorporate the detailed detector designs described in Section~\ref{sec:detect-a}
%Section~IV~A 
and the performance levels justified by R\&D as reviewed in Section~\ref{sec:detectrd}.
%Section~IV~B. 
This gives our estimates denoted S1* (See Figure 2 caption).  The inputs are described in more detail in \cite{ILCforESS}. 
For the nominal ILC program at 250~GeV, the Higgs
coupling to $b$ quarks is expected to be measured to 1.1\% accuracy and the
couplings to $W$ and $Z$ to 0.7\% accuracy. 
The full set  of  expected
uncertainties  is shown in Fig.~\ref{fig:ILCmodelindep}. 

In a manner similar to the estimates in \cite{Yellow}, a more optimistic scenario S2* is defined, assuming that detector performance can be improved with experience.   The precise scheme is described in \cite{ILCforESS}.   The S2* estimates are also shown in Fig.~\ref{fig:ILCmodelindep}.   The blue bars in the figure show the improvement in the errors when running at 500~GeV is also included.    The 
discovery of any anomaly at 250~GeV can be confirmed 
using additional reactions  such as $WW$-fusion production of the
Higgs boson.   Measurements at this
level can discover---and distinguish---models of new physics over a
wide space of possibilities, even for models in which the predicted new
particles are too heavy to be discovered at the
LHC~\cite{Barklow:2017suo}.

Figure~\ref{fig:ILCLHC} compares the ILC projections to those given in the
HL-LHC Yellow Report~\cite{Yellow} in their scenarios S1
and S2. The LHC projections include
model-dependent assumptions.  To assist the comparison,
these assumptions are imposed also in the ILC analyses. The 
uncertainties in the extracted Higgs couplings under these 
assumptions~\cite{ILCforESS} are shown as the  S1 and S2 values in the figure. The blue bars again show the effect of adding a data set at 
500~GeV, as described in \cite{ILCforESS}.

In addition to its decays predicted in the SM, the Higgs boson could have additional decays 
to particles with no SM gauge interactions.    These decays may
include invisible decays (\eg, to a pair of dark matter particles $\chi$)  or
partially invisible decays (\eg, to $b\bar b \chi \chi$).   The ILC
can robustly search for all types of exotic decays  to the 
part-per-mille level of branching fractions~\cite{Liu:2016zki}.

The ILC can also search for particles produced through electroweak
interactions, closing gaps that are left by searches at the LHC. These include searches for dark matter candidates and for the radion and dilaton of extra-dimensional models. An
important example is the Higgsino of supersymmetric models.   If the
mass  differences among Higgsinos are smaller than a few GeV---as predicted in currently allowed models---then Higgsinos
of 100~GeV mass would be produced copiously at the LHC, but this
production would not be registered by LHC triggers.  In the clean
environment 
of the ILC, even such difficult signatures as this 
would be discovered and the new particles 
studied with percent-level precision~\cite{Higgsino}.

ILC precision measurements of $\ee\to f\bar f$ processes at 250~GeV have a sensitivity to new
electroweak gauge bosons comparable to (and complementary with) 
direct searches at the LHC. Though the center of mass energy is only a little higher than that of LEP 2, the ILC will collect an event sample 1000 times greater, with detectors dramatically superior in their heavy flavor identification and other capabilities.  Polarisation plays a
key role since it allows  the electroweak couplings to be
disentangled, with particular 
sensitivity to right-handed couplings.   The reaction $\ee\to b\bar b$ 
is of special interest since it can receive corrections not only from new electroweak interactions but also from new physics 
that acts primarily on the Higgs and the heavy quark doublet $(t,b)$~\cite{eetobb1,eetobb2}. 

The ILC at 250~GeV can be the first step to the study of $\ee$
reactions at higher energy.   A linear $\ee$ collider is extendable in
energy by making the accelerator longer or by increasing the
acceleration gradient. Extensions to 500~GeV and 1~TeV were envisioned
in the ILC Technical Design Report~\cite{Behnke:2013xla}.   The aims
of 
this higher-energy program are discussed in detail in
\cite{ILCforESS}.
They include the
measurement of the top-quark mass with a precision of 40~MeV, 
measurements of the top-quark
electroweak couplings to the per-mille level, measurement of the
Higgs coupling to the top quark to 2\% accuracy, and measurement of
the triple-Higgs boson coupling to 10\%  accuracy.   Higher energy stages of the ILC will also extend searches for new particles with electroweak interactions and will give sensitivity to new $Z'$ bosons of mass 7--12~TeV. 
Eventually, the ILC tunnel could be the host for  very high
gradient electron accelerators reaching energies much higher than 1 TeV.   The 
ILC promises a long and bright future beyond its initial 250~GeV stage.

\vspace{-.4cm}

\section{\label{sec:collider}Collider}

\vspace{-.3cm}

%\todo{ 2 pages - Benno, Shin
%Summary of the ILC250 design (important to note the elements that are retained in first stage to accommodate future energy extensions) }

\begin{table}
\begin{tabular}{lccccc}
Quantity & Symbol & Unit & Initial &  \multicolumn{2}{c}{Upgrades} \\
\hline
Centre-of-mass energy & $\sqrt{s}$ & ${\mathrm{GeV}}$ & $250$ & $500$ & $1000$ \\
Luminosity & \multicolumn{2}{c}{${\mathcal{L}}$ $(10^{34}{\mathrm{cm^{-2}s^{-1}}}$})& $1.35$ & $1.8$ & $4.9$ \\
Repetition frequency &$f_{\mathrm{rep}}$ & ${\mathrm{Hz}}$  & $5$ & $5$ & $4$ \\
Bunches per pulse  &$n_{\mathrm{bunch}}$ & 1  & $1312$ & $1312$ & $2450$ \\
Bunch population  &$N_{\mathrm{e}}$ & $10^{10}$ &$2$ & $2$ & $1.74$ \\
Linac bunch interval & $\Delta t_{\mathrm{b}}$ & ${\mathrm{ns}}$ & $554$ & $554$ & $366$ \\
Beam current in pulse & $I_{\mathrm{pulse}}$ & ${\mathrm{mA}}$& $5.8$ & $5.8$ & $7.6$  \\
Beam pulse duration  & $t_{\mathrm{pulse}}$ & ${\mathrm{\mu s}}$ &$727$ & $727$ & $897$ \\
Average beam power  & $P_{\mathrm{ave}}$   & ${\mathrm{MW}}$ & $5.3$   &$10.5$  & $27.2$ \\  
Norm. hor. emitt. at IP & $\gamma\epsilon_{\mathrm{x}}$ & ${\mathrm{\mu m}}$& $5$ & $10$ & $10$  \\ 
Norm. vert. emitt. at IP & $\gamma\epsilon_{\mathrm{y}}$ & ${\mathrm{nm}}$ & $35$ & $35$ & $35$ \\ 
RMS hor. beam size at IP  & $\sigma^*_{\mathrm{x}}$ & ${\mathrm{nm}}$  & $516$ & $474$ & $335$ \\
RMS vert. beam size at IP &$\sigma^*_{\mathrm{y}}$ & ${\mathrm{nm}}$ & $7.7$  & $5.9$ & $2.7$ \\
Site AC power  & $P_{\mathrm{site}}$ &  ${\mathrm{MW}}$ & $129$ & $163$ & $300$ \\
Site length & $L_{\mathrm{site}}$ &  ${\mathrm{km}}$ & $20.5$ & $31$ & $40$ \\
\end{tabular}
\caption{Summary table of the ILC accelerator parameters in the initial $250\,{\mathrm{GeV}}$ staged configuration
and possible upgrades.
\label{tab:ilc-params}}
\vspace{-0.5cm}
\end{table}

The fundamental goal of the design of the ILC is to fulfill the
physics objectives outlined in this document  with high energy-efficiency.  In the design,
the overall power consumption of the accelerator complex during operation is limited to $129\,{\mathrm{MW}}$ at  $250\,{\mathrm{GeV}}$ and $300\,{\mathrm{MW}}$ at  $1\,{\mathrm{TeV}}$, which is comparable to the power consumption of CERN today.
% Stapnes at ALCW2018: 1.35TWh in 2012 -> 154MW on average in 2012
This is achieved by the use of SCRF technology for the main
accelerator, which offers a high RF-to-beam efficiency through the use
of superconducting cavities.  The cavities are operated at 
 $1.3\,{\mathrm{GHz}}$, where high-efficiency klystrons are commercially available.
At accelerating gradients of $31.5$ to $35\,{\mathrm{MV/m}}$, this
technology offers high overall efficiency and reasonable investment
costs, even considering the cryogenic infrastructure needed for the
operation 
at $2~\mathrm{K}$. Some relevant parameters are given in Table~\ref{tab:ilc-params}.

The underlying TESLA SCRF technology is mature, with a broad industrial
base throughout the world, and is in use at a number of free-electron-laser
facilities that are in operation (European XFEL at DESY), under construction (LCLS-II at SLAC),
 or in preparation (SHINE in Shanghai) in the three regions that have
 contributed to the ILC technology. In preparation for the ILC, Japan and
 the U.S.\ 
have founded a collaboration for further cost optimisation of the TESLA technology.
In recent years, new surface treatments during the cavity preparation
process, such as the so-called nitrogen infusion, have been 
developed at Fermilab and elsewhere.
These offer the prospect of  achieving  higher gradients and lower loss
rates than
assumed in the TDR, using a less expensive surface-preparation scheme.
This  would lead to a
 further cost reduction over
the current
 estimate.
 
 The design goal of energy efficiency fits well into the ``Green ILC'' concept \cite{GreenILC} that pursues a comprehensive approach to a sustainable laboratory.  Current European Research and Innovation programmes include
 efficiency studies for the ILC and other accelerators. 
A model is the recently inaugurated European Spallation Source ESS in Sweden, which followed the 4R strategy: Responsible, Renewable, Recyclable and Reliable.

When the Higgs boson was discovered in 2012 and the Japan
 Association of High Energy Physicists (JAHEP) made a proposal to host the ILC in Japan,
the Japanese ILC Strategy Council conducted a survey of possible sites
for the ILC in Japan, looking for  suitable geological conditions
 for a tunnel up to $50\,{\mathrm{km}}$ in length, and the possibility to establish a
 laboratory where several thousand international scientists could  work and live. 
The candidate site in the Kitakami region in
 northern Japan, close to the larger cities of Sendai and Morioka, 
was found to be the best option. 
The site offers a large, uniform granite formation, 
with no active seismic faults, that is well suited for tunnelling.
Even in the great Tohoku earthquake of 2011,  underground installations
in this rock formation were essentially unaffected. This  underlines
 the suitability of this candidate site. 

\begin{figure*}[tb]
 \begin{center}
 \includegraphics[width=\hsize]{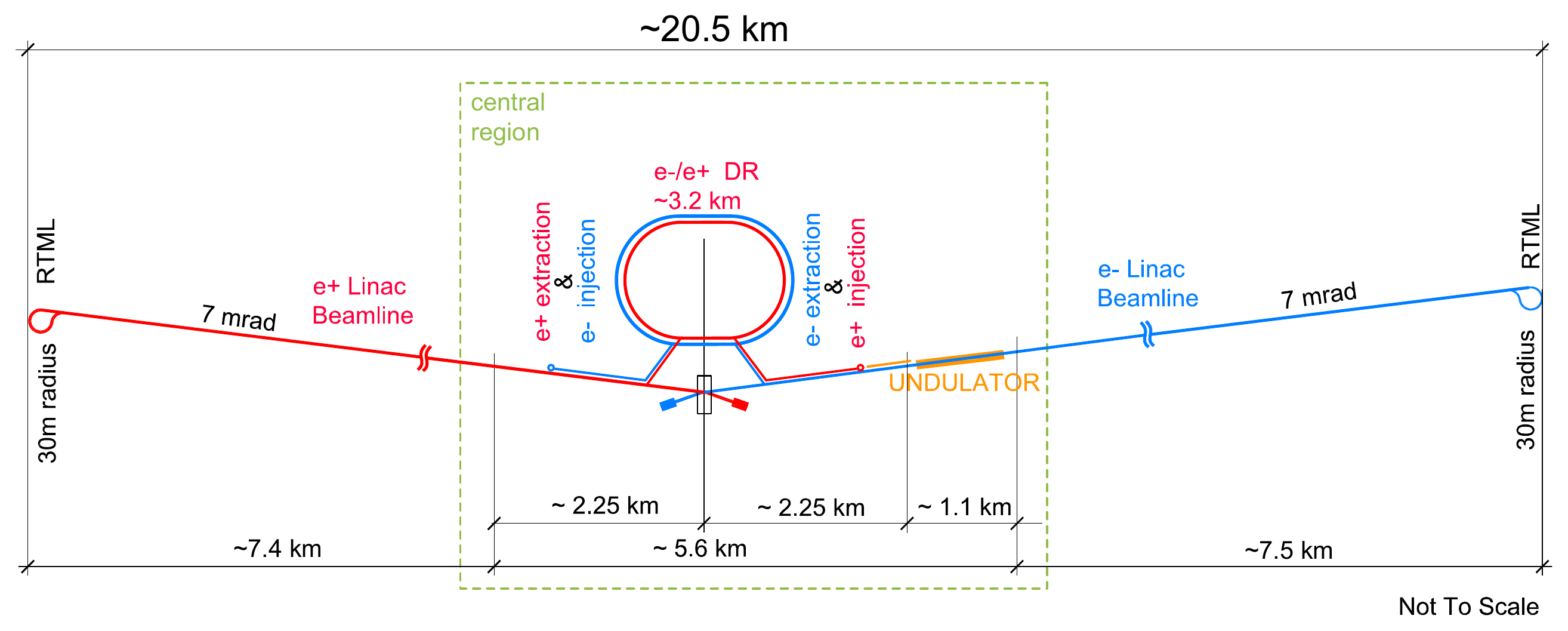}
\caption{Schematic layout of the ILC
 in the $250\,{\mathrm{GeV}}$ staged configuration.
\label{fig_ilc-schematic}}
 \end{center}
 \vspace{-0.7cm}
 \end{figure*}

Figure~\ref{fig_ilc-schematic} shows a schematic overview of the
initial-stage 
accelerator with its main subsystems.
The accelerator extends over $20.5\,{\mathrm{km}}$, with two main arms that are dominated by the electron and positron main linacs, at a $14\,{\mathrm{mrad}}$ crossing angle.

Electrons are produced by a polarised electron gun located in the
tunnel of the positron beam-delivery system. A Ti:sapphire laser
impinges on a photocathode with a strained GaAs/GaAsP superlattice
structure, which will provide  $90\,\%$ electron polarisation at the
source, resulting in more than $80\,\%$ polarisation at the interaction
point. The design is based on the electron source of the SLAC
Linear Collider (SLC). 

Two concepts for positron production are considered.
The baseline solution employs superconducting helical undulators at
the end of the electron main linac, producing polarised photons that
are converted to positrons
in a rotating target, with a $30\,\%$
 longitudinal polarisation. 
This positron-production scheme requires an operational electron linac
delivering a beam close to its nominal energy of
$125\,{\mathrm{GeV}}$, 
which is a complication for commissioning and operation. 
%The main technical challenges of this concept are the target and the photon dump. 
%In addition, the undulator photon flux rapidly falls for electron energies below $125\,{\mathrm{GeV}}$, which is a concern for commissioning and operation.  
An alternative design, the electron-driven source, utilises a dedicated S-band electron accelerator to provide a $3\,{\mathrm{GeV}}$ beam that is used to produce positrons by pair production.
% on a rotating titanium target. 
%Electrons are produced over a timespan of $66\,{\mathrm{ms}}$, reducing the target heat load and the necessary rotation speed.
%Technical challenges for this concept are radiation in the capture device and beam loading in the first accelerating cavities. 
This source might not provide positron polarisation,
but would have advantages for operation at lower electron beam energies and during commissioning.
% as its operation is independent  from the main linac electron beam, operation at lower electron beam energies is possible and commissioning can begin in parallel to electron source commissioning.
Both concepts are
likely to prove viable when the requisite engineering effort can be devoted to their design.
The current accelerator design is compatible with either option. 
A decision between the alternatives will be made before commencement of the detailed engineering design, based on their relative physics potential, costs, and technical maturity.

Electrons and positrons are injected at $5\,{\mathrm{GeV}}$ into the
centrally placed $3.2\,{\mathrm{km}}$-long damping-ring complex, where
their normalised emittance is reduced to $20\,{\mathrm{nm}}$
($4\,{\mathrm{\mu m}}$) 
in the vertical (horizontal) plane within $100\,{\mathrm{msec}}$. 
These emittance numbers are well in line with 
the performance of today's storage rings for advanced light sources.
To achieve the necessary damping time constant,
 the damping ring is equipped with $54$ superconducting wigglers. 

The damped beams are transported to the beginning of the main
accelerator by two low-emittance beam-transport lines. A two-stage
bunch compressor from $5$ and $15\,{\mathrm{GeV}}$ reduces the
longitudinal bunch length to $300\,{\mathrm{\mu m}}$ before the beams
are accelerated to $125\,{\mathrm{GeV}}$
 in the two main linacs.

The main linacs accelerate the beams in superconducting cavities made of niobium, operating at $1.3\,{\mathrm{GHz}}$ frequency and a temperature of $2.0\,{\mathrm{K}}$. 
Each cavity has $9$ cells and is $1.25\,{\mathrm{m}}$ long. 
The mean accelerating gradient will be $31.5$ to $35\,{\mathrm{MV/m}}$.
Cavities are mounted in $12\,{\mathrm{m}}$-long cryomodules that house $9$ cavities or $8$ cavities plus a quadrupole unit for beam focusing. 
The cryomodules provide cooling and thermal shielding and contain all
necessary pipes for fluid and gaseous helium at various temperatures. 
No separate helium transport line is necessary.

Cryomodules of this type have been in continuous operation since 2000 in the TESLA Test Facility (TTF, now FLASH), since 2016 at the FAST facility at Fermilab where the ILC specification of the 31.5 MeV/m beam acceleration gradient was demonstrated~\cite{Broemmelsiek:2018iqr},  and, since 2017, 97 of these cryomodules have been in operation at the European XFEL.
This proves their 
long-term stability. 
Cost and performance estimates for the ILC cryomodules are based on
the 
experience from these facilities, and thus can be regarded with high confidence. 

The RF power for the cavities is generated by commercially available $10\,{\mathrm{MW}}$ klystrons with an efficiency of $65\,\%$. 
The pulse modulators will use a new, modular and cost-effective semiconductor design \jim{developed} at SLAC, the MARX modulator.

The cryogenic design for the superconducting cavities is planned with six cryo plants for the
main linacs, each with a size similar to those operating at CERN (8
plants for the LHC), 
DESY (for HERA/ XFEL) and Fermilab (for the Tevatron).
Two smaller plants would supply the central region, including the preaccelerators of the sources and the damping rings. 

Finally, the beam-delivery system focuses the beams 
to the required size of $516\,{\mathrm{nm}} \times 7.7\,{\mathrm{nm}} $. 
A feedback system, which profits from the relatively long
 inter-bunch separation of $554\,{\mathrm{ns}}$, ensures the necessary beam stability. 
The necessary nano-beam technology and feedback control has been tested at the Accelerator Test Facility 2 (ATF-2) at KEK, where beam sizes of $41\,{\mathrm{nm}}$ have been demonstrated~\cite{Okugi:2017jji}; these correspond to the ILC design goal within $10\,\%$ after scaling for different beam energies.

The TDR baseline design assumed
 a centre-of-mass energy of $\sqrt{s}=500\,{\mathrm{GeV}}$, upgradeable to a final energy of $1\,{\mathrm{TeV}}$.
After the discovery of the Higgs boson in 2012, interest 
grew for an accelerator operating as a ``Higgs factory'' at
 $\sqrt{s}=250\,{\mathrm{GeV}}$, slightly above the maximum for $Zh$ production. 
The design for a $250\,{\mathrm{GeV}}$ version of the 
ILC has recently been presented in a  staging
 report by the LCC directorate~\cite{Evans:2017rvt} and was endorsed by ICFA.

This staged version of the ILC  
% (in the preferred ``option A'') 
would have two main linac tunnels about half the length of the
$500\,{\mathrm{GeV}}$ TDR design    ($6\,{\mathrm{km}}$  instead of $11\,{\mathrm{km}}$).
Other systems, in particular the beam-delivery system and the 
main dumps, would retain the dimensions of the TDR design.
Then the ILC250 could be upgraded to energies of $500\,{\mathrm{GeV}}$
or even $1\,{\mathrm{TeV}}$ with a reasonable effort, without
extensive
 modifications to the central region. 
Recent studies of rock vibrations from tunnel excavation in a similar
geology indicate that the necessary additional main linac tunnels
could be largely constructed during ILC operation, so that an energy
upgrade could be realised with an interruption in data taking of only
about 2~years, 
compatible with a smooth 
continuation of the physics programme.

Another upgrade option, which could come before or after an energy upgrade, is a luminosity upgrade. 
Doubling the luminosity by doubling the number of bunches per pulse to $2625$ at a reduced bunch separation of $366\,{\mathrm{ns}}$ would require $50\,\%$ more klystrons and modulators and an increased cryogenic capacity. 
The damping rings would also permit an increase of the pulse
 repetition rate from $5$ to $10\,{\mathrm{Hz}}$ at $250\,{\mathrm{GeV}}$ .  
This would require a significant increase in cryogenic capacity, 
or running at a reduced gradient after an energy upgrade.
% Both paths to a luminosity increase would pose large  stress for the positron source target and probably require its redesign, which would be a significant engineering challenge, but not a cost driver.
The projections for the physics potential 
of the ILC250 are based on a total integrated 
luminosity of $2\,{\mathrm{ab}}^{-1}$, which assumes at least one luminosity upgrade.

%The total project cost for the construction of the
%$250\,{\mathrm{GeV}}$ accelerator is estimated to be
%$5.3\,{\mathrm{BILCU}}$ plus
% $17\,{\mathrm{Mh}}$ of work, where the   ILCU (ILC Currency Unit) 
%is defined as equal to  a 2012 US\$.
%These numbers include the cost for civil engineering and the
%laboratory. Costs for land acquisition, R\&D
% costs before the construction starts, and costs for the detectors are
% not included.
%The  cost is calculated as an estimate of the 
%median project cost, without contingency or management reserve.
%The cost premium to cover the project cost with $85\,\%$ instead of
%$50\,\%$ confidence level (loosely speaking, the $1\,\sigma$ 
%uncertainty of the cost estimate) has been estimated
% to be $23\,\%$ of the estimated cost.

%The construction of the accelerator will require 9 years after ground
%breaking, preceded by a 4-year preparation phase.

\vspace{-.4cm}

\section{\label{sec:detect} Detectors}

\vspace{-.3cm}

%\todo{ 2 pages - Ties, Andy}

%\todo{introductory paragraph is needed to connect with the R\&D section and giving continuity to the whole programme: detector designed being the consequence of the Machine conditions, physics requirements and detector R\&D progress.}

The detector concepts proposed for the ILC have been developed over
the past 15 years in a strong international effort. They reflect the
requirements placed on the detectors from the science, and have folded
in the constraints from the design of the machine, in particular the
special properties of the 
interaction region.  They incorporate 
 the results of  the R\&D effort described below.

\vspace{-.4cm}

\subsection{\label{sec:detect-a}The full detector systems, ILD and SiD}

\vspace{-.3cm}

The main guiding principles for the full detector systems are:

\begin{itemize}
 \item The detector must have excellent track momentum
   resolution, of about $\delta(1/p)=2 \times 10^{-5} \mathrm{GeV}^{-1}$. 
The benchmark  here is the analysis 
of the di-lepton mass in the process $\ee \to HZ \to H \ell^+
\ell^-$. This reaction allows the reconstruction of the 
Higgs mass,  independently of its decay mode,  via the 
reconstruction of the lepton recoil momentum. The Higgs boson mass is
important by itself, but it is also a crucial input in the precise SM
prediction of the Higgs boson properties.  Stringent momentum resolution requirements must be reached
to meet the mass resolution goal. 
\item Many physics measurements depend on the flavor  identification of heavy
  quarks
and leptons.  For this, very powerful vertex detectors 
are needed. Both for the known Higgs boson and, typically, for 
extended Higgs particles, the most prominent decays are to
third-generation
species.   Many other physics processes also lead to 
complex  final states containing  bottom or charm quarks.
A superb  vertex detector  is needed to reconstruct these 
long-lived particles 
precisely and with high efficiency.
 For example, the position of the reconstructed secondary
 vertex
 should be found with a precision of better than $4\,\mu$m.
\item The momenta of the full set of final-state particles are
 best reconstructed with the 
Particle Flow Algorithm (PFA).
 This  technique combines 
the information from the tracking systems and from the 
calorimetric systems to reconstruct the 
energy and the direction of all charged and 
neutral particles in the event. To minimise overlaps between 
neighboring particles, and to 
maximize the probability to 
correctly combine tracking and calorimeter information, 
excellent calorimeters with very high granularity  are needed.
The agreed-upon goal is  a jet energy resolution of $3 \%$ -- an
improvement of about a 
factor of two over the LHC detectors. 
\item Many physics signatures predict some undetectable particles
which escape from the detector. These  can only be reconstructed by 
measuring the missing energy and 3-momentum  in the event. This requires 
that the detector is as hermetic as possible. 
Particular care must be given to the region at small angles surrounding the 
beampipe. %Section~\ref{sec:detect-a}
\end{itemize}

\begin{figure}[tb]
 \begin{center}
 \includegraphics[width=0.9\hsize]{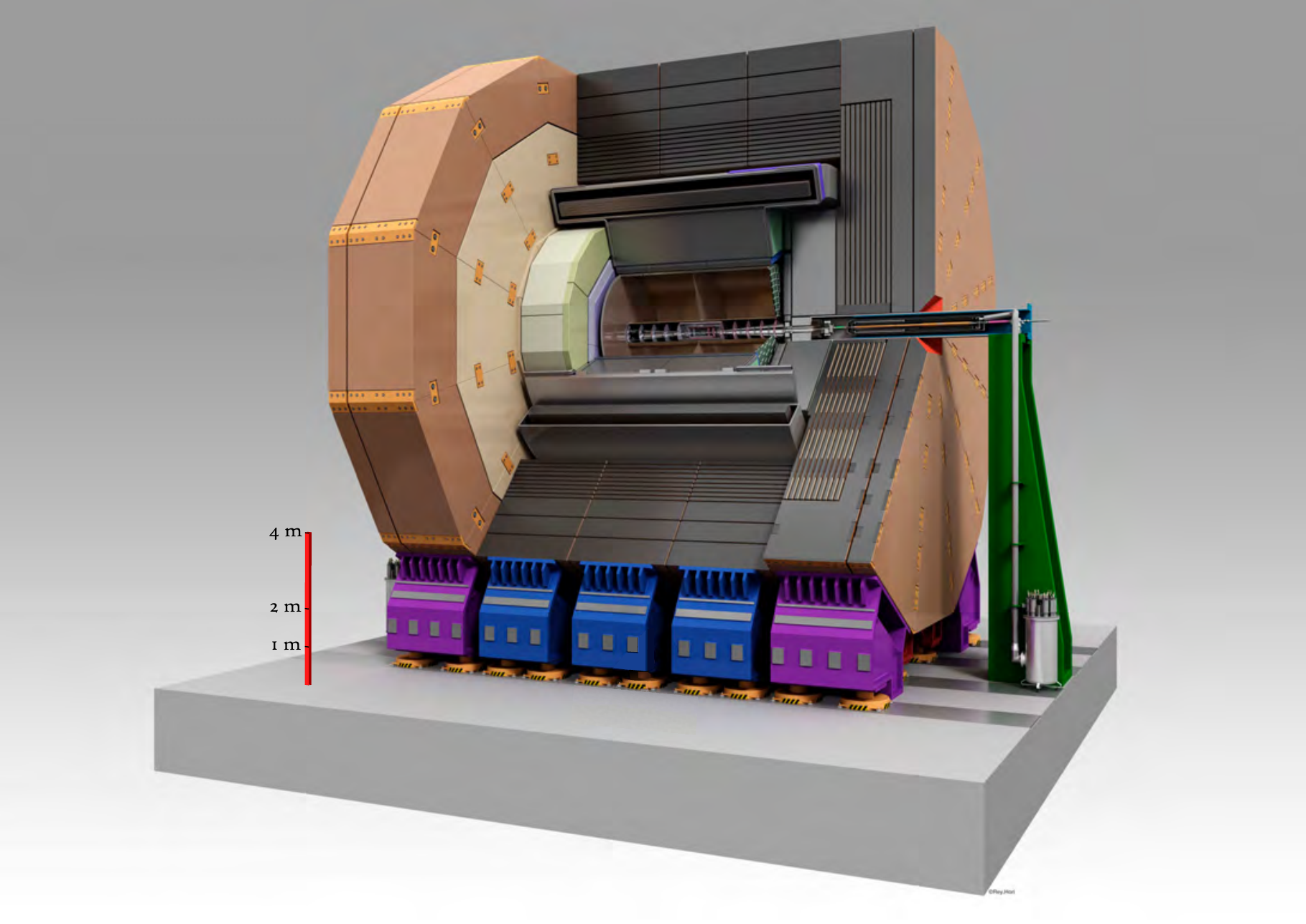}
\caption{The ILD detector concept.
\label{fig_ild}}
 \end{center}
 \vspace{-0.7cm}
 \end{figure}
 
Compared to the last large-scale detector project in particle physics,
the construction and upgrade of the LHC detectors, the 
emphasis for linear collider detectors
 is shifted towards ultimate precision. This requires detector technologies
with new levels of performance.  It also requires the  minimisation of
passive material in the detector at an unprecedented level, with strict
management and control of services and, in particular,
thermal management of the detector.  As a benchmark,  the total
material in front of the electromagnetic calorimeter should not exceed
a few percent of a radiation length. 
This is possible due to the relatively low levels
of radiation, compared to the LHC, for example.
Significant technological R\&D was needed to
demonstrate the feasibility of this goal.

\begin{figure}[tb]
 \begin{center}
 \includegraphics[width=0.9\hsize]{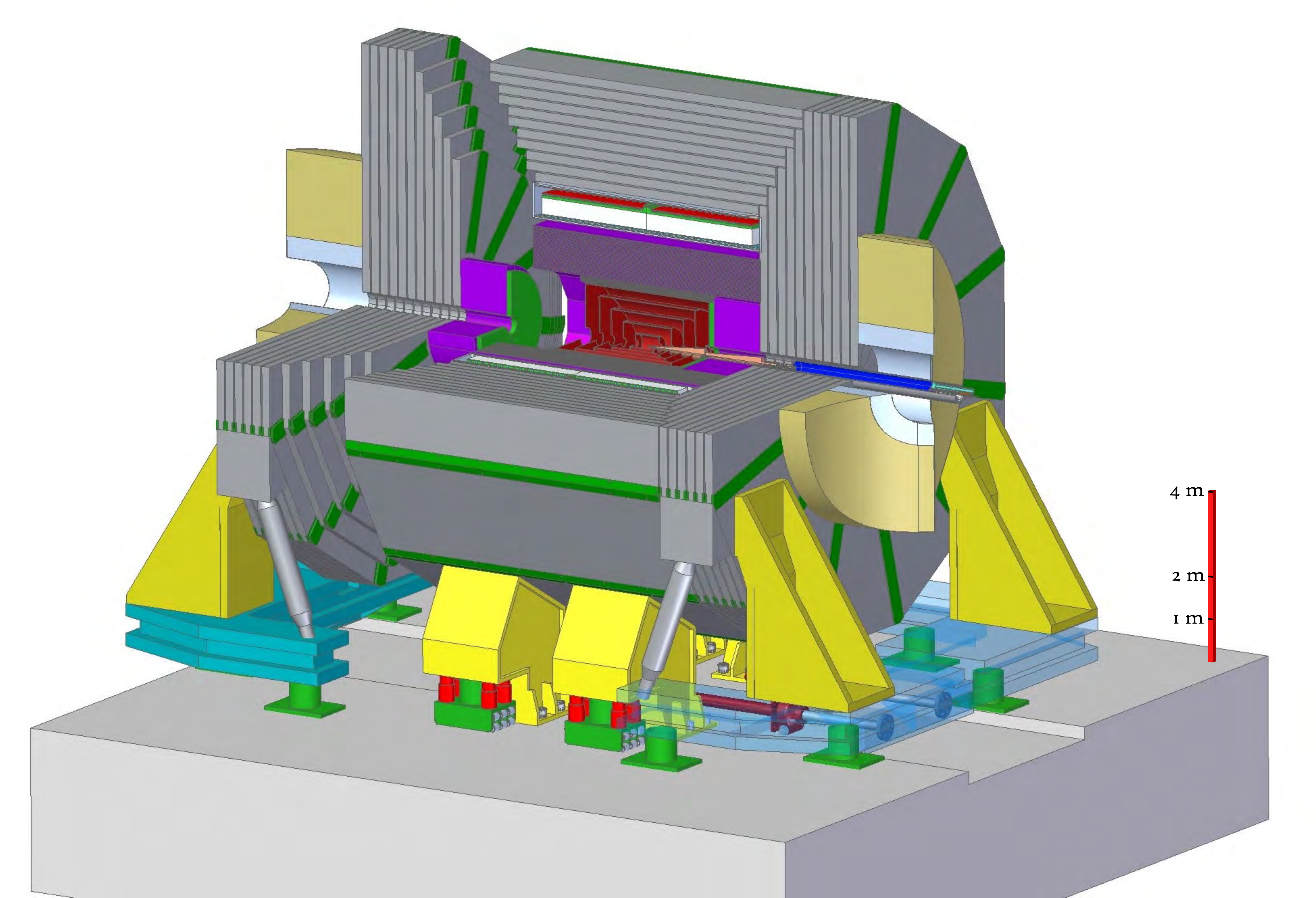}
\caption{The SiD detector concept.
\label{fig_sid}}
 \end{center}
 \vspace{-0.7cm}
 \end{figure}

Over the last decade, two detector concepts have emerged from the
discussions and studies in the community. Both are based on the assumption that
the 
particle-flow technique will  play a central role in the event
reconstruction. Both therefore have highly granular calorimeters
placed inside the solenoid coil
and excellent trackers and vertexing systems. The two approaches
differ in the choice of tracker technology, and in the approaches taken
to maximise the overall precision of the event reconstruction. ILD
(\Fig{fig_ild}) has chosen a gaseous central tracker, a time
projection chamber, combined with silicon detectors inside and outside
the TPC. SiD (\Fig{fig_sid}) relies on an all-silicon solution,
similar to the
 LHC detectors, although with
much thinner silicon layers. ILD would optimise the particle-flow
resolution by making the detector large, thus separating charged and
neutral particles. SiD keeps the detector more compact, and
compensates by using a higher central magnetic field. Both approaches
have demonstrated excellent performance through prototyping
 and simulation, meeting or even exceeding the requirements. 

The ILC infrastructure has been designed to allow for two detectors,
operated in a so-called push-pull mode. The detectors are mounted on
movable platforms, which can be moved relatively quickly in and out of
the beam. The goal is to exchange the detectors in the IP and be ready
to take data within 
a day or two. 

This baseline design with two detectors
has distinct scientific advantages over a
one detector arrangement.
The push-pull design is much less expensive than 
that with two separate interaction points. 
The scientific advantages arise from the
complementarity of the detectors, the competition between detector
teams, the opportunity for independent cross-checks of new results,
and
 the likely larger community of participants in the scientific program. 

For both detector concepts, communities have self-organised and
pre-collaborations have formed. Over the last ten years, 
 these organisations have pushed both concepts to a remarkable level of
maturity. In close interaction with the different groups
performing detector R\&D from around the world, they have demonstrated the
feasibility of building  and operating such high-precision 
detectors. 

European groups have played a central role in these efforts. The ILD
concept group is formed from some 70 groups from around the world,
with more than half coming from Europe. The SiD collaboration has a
strong basis in the Americas, but also relies on significant
participation from European groups. Major contributions to the
development of all sub-systems have come from Europe. Significant
technological breakthroughs, for example in the area of highly granular
calorimeters, are
 strongly driven by European groups. 

An important aspect of the detector concept work has been
the integration of the detector into the collider and into the
proposed site. The location of the experiment in an earthquake-prone
area poses challenges which have been addressed through R\&D on
detector stability, support and service. The scheme to operate two
detectors in one interaction region required
significant engineering work to demonstrate its
feasibility. With strong support from particle physics laboratories in
Europe, in  particular DESY and CERN, many of the most relevant
questions were answered and the feasibility
 of the approach demonstrated, at least in principle. 

%%%%%%%%%%%%%%

\vspace{-.4cm}

\subsection{\label{sec:detectrd} Detector R\&D}

\vspace{-.3cm}

%\todo{ 1 page - Marc.}
%\todo{At present the R\&D section is copied from the original text by
%Marc from Doc2. It needs to be better adapted to Doc1 and a more
%natural merging/transition to the detectors section. Industrial
%participation, interest and motivation should also be implemented.}
%\todo{I have polished the English in this section and made a few
%connections to the Detectors section, but this is not as extensive a
%revision as called for in the previous note. -- MEP}

The physics demands for high precision challenge
the ILC detector designs.  Optimal 
trade-offs between
granularity, material, speed and power, and 
resolution were needed to achieve
the performance parameters  discussed
in Section~\ref{sec:detect},
an order of magnitude
improvement in state-of-the-art.
Intensive R\&D was needed to realise this performance,
reliably and at minimal cost,
on the subsystem level, and  within the complete,
integrated detector system \cite{RDliaision}.

%The ILC accelerator will produce well-characterized 
%initial beam particles and relatively mild
%running conditions without strong pressure for  radiation-tolerance
%and high readout speed.   However, the demands from the physics to
%achieve high precision are very strong. It has been a  challenge,
%then, to  use the ILC running conditions optimally, with an
%appropriate  trade-off
%between very demanding goals for resolution, granularity, and material
%budget
%on the one hand and acceptable speed and low power on the
%other hand.
%In addition to these demands, the ILC data-taking is expected to
%operate without a trigger, even while the detector operation exploits
%the machine duty cycle and bunch spacing to minimise the power
%consumption.
%The individual detector elements need to achieve their high
%performance -- in most cases, an order of magnitude improvement over
%the state of the art -- while integrating with one another at the
%system level.  In several cases, the individual performances targeted 
%by the R\&D might  have been considered as nearly achieved outside
% of the ILC programme, but intensive R\&D was needed to realise this
% performance within the complete detector system.   The goals of the
% ILC R\&D was then to prototype each sub-system to provide a realistic 
%design for the full detector with reliable performance and minimal
%cost. 

The use of the Particle Flow Algorithm (PFA) for reconstruction
of final-state particles based on both tracking and calorimetric information
required an integrated approach.  
Once optimized, the results could be applied
to the 
realistic Monte Carlo simulations of
physics performance discussed in Section~\ref{sec:phys}.
Several variants of
calorimeter and tracking subsystems were
developed for studies in test beams,
sometimes within a $2\,$T magnetic field.
These enabled studies that optimized performance with respect to cost constraints.

Tracking and vertex detector development
  was driven by pixellated, low-material
 components with excellent momentum resolution and displaced
 vertex characterisation, including vertex charge, performances
typically exceeding existing experiments  by an order of magnitude.

%The development of tracking and vertexing
% devices was driven by the need for pixellated low material budget
% components allowing excellent momentum resolution and displaced
% vertex characterisation, including vertex charge, with performances
%typically  exceeding  by an order of magnitude those
% of existing 
%experiments.

Two main tracker alternatives  were investigated:
a TPC and silicon sensors, possibly pixelated. TPC R\&D 
addressed mainly the single-point
resolution and ion-feedback mitigation with different micro-pattern
read-out systems (MicroMegas, GEM, $\ldots$), showing
performance goals are reached, with an end-cap material budget of 
less than $30 \%\,X_0$. Silicon sensor R\&D aimed at reducing the material budget; 
targeted momentum resolution is achieved with a limited number of layers.
ATLAS and CMS tracker upgrade R\&D contributed,
although ILC silicon tracking layers
are much thinner with somewhat different solutions.
A large-area pixelated tracker may improve
performance over silicon-strips in dense jet environments.  

%Two alternative approaches were investigated for the main tracker, one
%based on a TPC and one exploiting silicon sensors, possibly
%pixelated. The R\&D for the TPC addressed mainly the single-point
%resolution and ion-feedback mitigation with different micro-pattern
%read-out systems (MicroMegas, GEM, $\ldots$) and showed that the
%performance goals can indeed be reached, with a material budget of the
%end-caps not exceeding $30 \%\,X_0$. The R\&D for the
%silicon detectors concentrated on the material budget, showing that
%the targeted momentum resolution could be reached despite the
%restriction
%on the 
%number of detector layers allowed by mitigating multiple-scattering
%effects. These efforts have benefited from the R\&D that
%has been conducted for the tracker upgrades for ATLAS and CMS,
%both of which rely on all-silicon tracking systems. 
%Still, the tracking layers 
%envisaged for the ILC are much thinner 
%than those at the LHC as a consequence of the different beam
%conditions, leading to somewhat different solutions.
%It was also shown that, while the performance was more than adequate
%in terms of momentum resolution, the tracking in dense jet
%environments could be improved by replacing the silicon-strip sensors
%with   a large-area pixelated tracker. 

Vertex detector R\&D explored several
thin, highly-granular pixel technologies (CMOS, DEPFET, FPCCD, SoI,
$\ldots$) that offer the projected spatial resolution and material
budget. Intensive efforts focussed on read-out systems that 
handle the beam-related background hit density.  
The performance depends on material technology and read-out
architecture.   Double-sided layers were also investigated
establishing feasibility near an $\ee$
 interaction point.

%The R\&D for the vertex detector explored the potential of several
%thin, highly granular pixel technologies (CMOS, DEPFET, FPCCD, SoI,
%$\ldots$) which could offer the projected spatial resolution and material
%budget. Intensive efforts were invested in read-out systems that could
%cope with the hit density induced by the beam-related background.  The
%final performances depends both on the material technology and on the read-out
%architecture.   The concept of double-sided layers was also investigated
%with some technologies and established up to the level of being
%operated near an $\ee$
% interaction point.

PFA requirements lead to very compact, highly-granular
calorimetric technologies, including
low-power
read-out micro-circuits
with power pulsing.
The CALICE Collaboration studied the major issues
for both electromagnetic (ECAL) and  hadron calorimeters (HCAL).
ECAL R\&D concentrated on
optimised and cost-effective sensor systems, designs of 
low-power, pulsed, integrated readout electronics and effective
thermal management and calibration strategy, and  a
mechanical concept combining high stability with minimal passive material
zones. 
A SiW-based full-size prototype was constructed and tested
extensively on particle beams. 
A cost-effective scintillator/photo-sensor solution
was also tested.

%The development of calorimetric detectors was driven by the PFA
%requirements
%for 
%very compact, highly granular detection technologies, while at the
%same time incorporating 
%read-out micro-circuits
%with very low average power,
%including power pulsing. A dedicated study of the major issues
%in calorimeter design was carried out by 
%the CALICE collaboration, a consortium composed of 
%more than 300 members coming from more than 50 institutes.

%The R\&D for the electromagnetic calorimeter (EM) concentrated on
%optimised and cost-effective sensor systems, on the designs of a
%low-power, pulsed, integrated readout electronics and an effective
%thermal management and calibration strategy, and on including a
%mechanical concept that  combines high stability with minimal dead
%zones. A SiW-based full-size prototype was constructed and tested
%extensively on particle beams. The development of a more
%cost-effective technological solution, based on a scintillator and
%photo-multiplier matrix was
% also realised; some aspects of its 
%performances were comparable to those of the SiW concept.

HCAL prototyping emphasized
efficient and precise neutral hadron shower reconstruction. Two
options developed with 
 stainless steel conversion material included scintillator tiles with
silicon photo-sensors read out with analog electronics, and more
highly-segmented
RPCs with one or two bit signal encoding.

%The hadron calorimeter (HCAL) prototyping was governed by the need for
%an efficient and precise reconstruction of neutral hadron
%showers. Two
%read-out options were developed using 
% stainless steel as the conversion material, one combining scintillator tiles with
%silicon photo-sensors read out with analog electronics, and an
%alternative approach based on gaseous devices
% (\eg, RPCs) with higher segmentation but with 
%signal encoding of one or two bits only.

Test-beam campaigns combining various ECAL and HCAL options 
demonstrate the relative merits, including PFA processing.
The energy and topology resolution requirements have been
demonstrated, including in power-pulsing operation.

%The relative merits of the different ECAL and HCAL options have been
%evaluated using combined test-beam campaigns with common data.  Data
%has been processed with PFA software that was developed and assessed
%at the same time. These experiments demonstrated the
% possibility of achieving  the targeted energy and
%topology resolutions even when operating in power-pulsed mode.

Very forward calorimeter technologies with robust
electron and photon detection for luminosity
and operations measurements show satisfactory
performance with $1\,$MGy tolerance.
Tungsten absorbers coupled with alternating GaAs sensor planes
included fast feedback for beam tuning.

%Substantial effort was invested in developing technological solutions
%for the very forward calorimeters designed for robust electron and
%photon measurements required for integrated-luminosity measurements and
%bunch-to-bunch machine-parameter monitoring. Satisfactory
%performances, including a radiation tolerance to $1\,$MGy, were obtained
%with tungsten absorber layers alternating with GaAs sensor planes read
%out with dedicated electronics featuring a dual-gain 
%charge amplifier providing fast feedback for beam tuning.

%A more detailed description of the R\&D program for the linear collider
%detectors can be found in the reference document \cite{RDliaision}.

%%%%%%%%%%%%%%

\vspace{-.4cm}

\section{\label{sec:soft}Software and Computing}

\vspace{-.3cm}
%\todo{ [1 page - Frank G. and Akiya
%Description of ILC software and computing requirements] }

It will only be possible to meet
the physics goals of the ILC programme if the excellent detector resolution
of the two proposed ILC detector concepts described above
 is complemented with powerful
and sophisticated algorithms for event reconstruction and data analysis.
For over a decade, the ILC community has developed and improved its
 software ecosystem \emph{iLCSoft}~\cite{bib:ilcsoft}, which
is based on the event data model LCIO~\cite{bib:lcio},
and the generic detector description toolkit DD4hep~\cite{bib:dd4hep}. 
The iLCSoft tools are used by both ILC detector concepts and also  by CLIC.
From the start, a strong emphasis has been placed on 
 developing flexible and generic tools that can easily be applied
to other experiments or new detector concepts. 
This approach of developing common tools wherever possible
has helped considerably in leveraging the limited manpower
 and putting the focus on algorithm development that
is crucial for the physics performance. 

\begin{figure}
\begin{center}
\includegraphics[width=0.85\hsize]{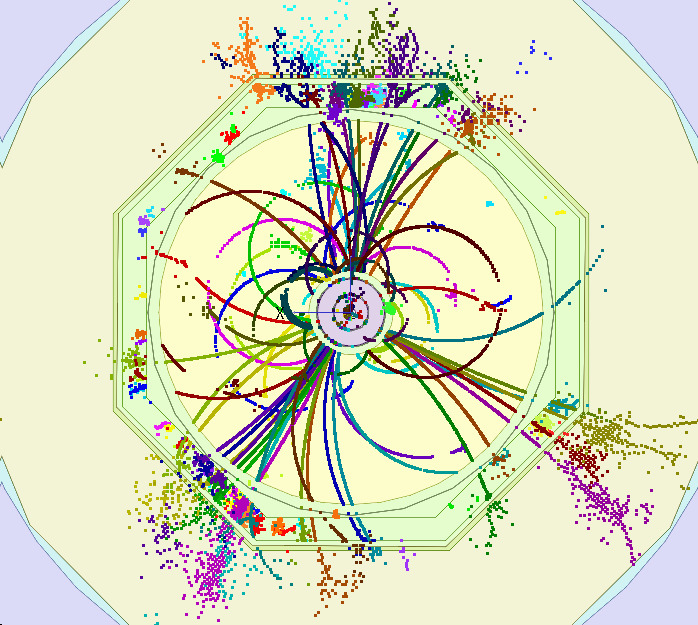}
\end{center}
\caption{Fully simulated and reconstructed $t\bar t$-event in the ILD
  detector, showing the individually reconstructed neutral and charged
  particles. The colour code presents the
  particle flow algorithm reconstruction without
  reference to the Monte Carlo generator
  information.}
\label{fig:ttbarevent}
\vspace{-0.7cm}
\end{figure}

A development of particular importance is the refinement of the 
PFA technique  that aims to identify
and reconstruct
 every individual particle created in the event
 in order to choose the best possible subdetector measurement for every particle. 
An example of individual particles reconstructed using PFA 
 in a $t\bar t$-event is shown in \Fig{fig:ttbarevent}.
%%%%%%%%%%%%%%%%
%%%%%%%%%%%%%%%%%%%%%%%%%%%%%%%%%%%%%%%%%%%%%%%%%%%%%%%%%%%%%%

Both detector concept groups have invested considerable effort into 
making their full-simulation models as realistic as possible.
Starting from a precise description of the actual 
detector technology, passive material, gaps and imperfections have been added.
Care has been taken to include realistic services such as cables and 
cooling pipes, in particular in the tracking region where
the material budget has a direct impact on the detector performance.
These simulation models have been used for large-scale Monte Carlo
production
 and physics analyses for the TDR and more recent detector optimisation
campaigns. Based on these studies, a 
realistic understanding of the expected detector performance and the physics
reach of the ILC for both detector concepts has been
achieved.

The development of \emph{iLCSoft} has been 
 a truly international activity, in which European groups, in particular DESY and
CERN, have played a leading role.  They 
will expand efforts if the ILC is approved.  The next stage
will 
focus on adapting the software tools for
 modern hardware architectures and continue to
improve the computing and physics performance of the algorithms.

An initial computing concept for the ILC, including a first estimate of the required resources, has been developed by the LCC Software and Computing Group~\cite{bib:lcc_computing}.
This concept follows in general terms that of the LHC experiments and
Belle~II,  with a strong on-site computing center complemented by large
Grid-based computing resources distributed around the world. Due to the much lower event rates at the ILC compared to the LHC, 
the detectors will run in an un-triggered mode in which  data from every bunch crossing will be recorded. At the detector site,  only limited computing
resources are required for online monitoring, QA, and data-buffering for a few
days.
 Prompt reconstruction, event building, and filtering of the interesting collisions
will be performed at the main ILC campus.
A few percent of the data will be distributed to major participating Grid sites
in the world for further skimming and final redistribution for physics
analysis. A copy of the raw data from all bunch crossings will be kept
to allow
 for future searches for new exotic signatures. 
Based on detailed physics and background simulations,
 the total raw data rate estimate of the ILC is $\sim$1.5 GB/s.
The total estimated storage needs will be a few tens of PB/y.
The computing power needed for simulation, reconstruction, and analysis will be a few hundred kHepSpec06.
Given that these numbers are already smaller than what is now
needed by the LHC experiments, and given an expected annual increase
of 15\% and 20\%, respectively, for storage and CPU
at flat budget, the overall computing costs for the ILC
will be more than an order of magnitude smaller than those for the LHC.

\vspace{-.4cm}

\section{\label{sec:discuss}Discussion and Summary}

\vspace{-.3cm}

%\todo{ 1 page - Keisuke, Jim and Juan
%Discussion of HEP community interest and support, political progress, plan for international realization, weight on ILC250 for searches and discovery potential, opportunity for young generations, maturity of the technology -detectors and accelerator-, reliability on cost estimates, etc..}

The ILC has a mature technical design
that is
 ready for construction. The ILC will start as a Higgs boson factory
 (ILC250).  Here the clean operating environment, low backgrounds, and
adjustable beam energies and polarisations will allow
model-independent 
measurements of the Higgs boson's mass and  $CP$ properties and of its
absolute couplings to SM fermions and gauge bosons, 
most of them to better than 1\% precision.  These measurements will
discriminate between the SM and many different BSM models.
The ILC will be sensitive to invisible and other exotic Higgs decays,
accessing additional new physics models including models of Dark
Matter.   The ILC polarised beams offer additional precision tests of
the SM, in particular for the electroweak couplings of right-handed
fermions, 
which are largely unconstrained today.

The ILC can be extended to higher energies in possible future
upgrades, up to 500 GeV and 1 TeV.  In these later stages, 
the ILC will give access to the properties of the top quark,
 including the top-quark Yukawa coupling, and  to the Higgs self-coupling.
Above  the top-quark production threshold,  the ILC will be
 a precision top-quark factory. Throughout its energy evolution,  the
 ILC will be able to produce pairs of new BSM particles of mass 
up to half its centre-of-mass energy and to  provide
 sensitivity to new force particles $Z'$ well beyond the direct search reach of the LHC.

Since no new particles beyond the SM have been 
discovered at the LHC, the search for new physics through
high-precision studies at the electroweak scale, particularly the Higgs boson and  the top quark, has
become
 urgent and compelling.  These studies strike at the heart of the
 mysteries of the SM in a way that is orthogonal to 
 direct searches for new particles.
 As discussed in Section~II, the ILC capabilities 
for precision tests
will be qualitatively superior  to those of the high-luminosity LHC. 
 This makes the ILC a powerful complement to  future LHC
 particle searches, with the ability to discover the new interactions that underlie the
 SM. 
 
The goal of a precise understanding of the Higgs boson is attractive
in its own right, readily communicated to our scientific
colleagues in other disciplines, as well as the general public.  Together
with this goal, the ILC provides a fully formed project proposal
with a  cost estimate
 similar to that of the LHC, a moderate time scale, and well tested
 technologies for its 
detector and accelerator designs. 

Future circular $\ee$ colliders have been proposed as an alternative 
method for precision Higgs boson studies.  These have the potential to
deliver higher luminosity at energies up to about 300~GeV.  However,
the ILC, operated as a Higgs factory, can take advantage of  beam 
polarisation to achieve similar physics
performance~\cite{Barklow:2017suo}.
More importantly, the straightforward energy
upgrade path of the ILC
makes the Higgs factory stage  only
the first phase of its potential for exploration. 

As emphasised in the previous sections, the ILC proposal
is supported by extensive R\&D and prototyping, both for the
accelerator and for the detectors.  For the accelerator,  
the successful construction and
operation 
of the European XFEL at DESY gives
us confidence both in the high reliability of the basic
technology and in the reliability of its performance and cost in 
industrial realization.   For the detector, an extensive course of
prototyping underlies our estimates of full-detector performance 
and cost.  Some specific optimizations and technological choices remain.
But the ILC is now ready to move forward to construction. 

 The ILC TDR
cost has been 
 rescaled for ILC250~\cite{Evans:2017rvt} and has recently been futher
 re-evaluated incorporating items specific to Japanese construction
 and accounting.  The current quoted cost estimate 
of the ILC250 is shown in Appendix A. 
 This cost has  been scrutinised in a number of
studies, most recently by a working group of the Japanese MEXT ministry, as described below.  Here too, the ILC is ready to move forward.

A strong community of universities and laboratories world-wide is
ready to realise the ILC, to develop its detectors, and to exploit its
physics opportunities. 
 The ILC Technical Design Report was signed by
2400 scientists from 48 countries and 392 institutes and university
groups,
 as described in Appendix B.  This community continues
 to prepare for the scientific program and
will expand its efforts once the ILC is launched as a project.
 
The ILC R\&D program and the construction of the FELs based on
SCRF in Europe, the US, and Asia, has opened strong links between the ILC
community 
and industry.  Very productive
 networking and communication has been established between industry
 representatives and scientists. Since 2016,
 all linear collider conferences have included one-day mini-workshops
 to show and promote industrial opportunities. 
These industrial mini-workshops have been well attended
 with growing  interest and participation from individual companies and from the industrial associations of several key countries.
 
On the political side, broad interest for the ILC in Japan has been
steadily growing.  The plan for  hosting the ILC
in Japan  is being promoted by political entities, at
the Japanese Diet and at the provincial levels, by a large industrial 
consortium (AAA), and by representatives of the particle physics
community (JAHEP).  Since 2013, the ILC project has been 
examined extensively by the MEXT ministry within a
cautious official procedure, in which  minimising risks is of
prime importance.  MEXT's ILC Advisory Panel released its
report~\cite{AdvPanel} on July 4, 2018.  This report summarises the
studies of the several working groups (WG) that
reviewed 
a broad range of aspects of the ILC.  The most recent studies include
a specific review of the scientific merit and the technical design for the ILC250. 
The  Physics WG scrutinised the scientific merit of the ILC250,
leading to their strong and positive statement on the importance of
the ILC250 to 
measure precisely the couplings of the Higgs boson \cite{AdvPanel}.
%\textit{If any coupling(s) is measured to be different from the Standard Model prediction, a particle-by-particle pattern of the deviation will elucidate the nature of new physics, suggesting a future direction of elementary particle physics. Mysteries in the Standard-Model such as the nature of dark matter and compositeness of the Higgs boson may also be clarified with this measurement}.
The TDR WG reviewed issues addressed in the Technical Design Report
and the ILC250 design, including the  cost estimate and technical feasibility.  
Other working groups of the MEXT review commented on manpower needs, 
organisational aspects, and the experience of previous large projects.
The report of the ILC Advisory Panel was followed by the beginning of
deliberations in a committee and technical working group 
established by the Science Council of Japan (SCJ).  
Another independent committee (ILC Liaison Council),
led by leaders of the Liberal Democratic 
Party, the majority party in the Diet,  has now 
convened to encourage the national government to proceed with the ILC.

It is an important aspect of the discussions of ILC in Japan that the
ILC is seen as a global project that will foster exchange between Japan
and other nations.   Thus, the  
scientific interest and political engagement of partner countries is a
major 
concern for the Japanese authorities.  For example, Japan has now
begun efforts to 
secure US partnership in the ILC.  The US Department of Energy Under
Secretary for Science recently visited Japan; he attended meetings
with political leaders promoting the ILC, and with the 
leadership of  KEK,
and stated the US would look forward
to a dialog on an ILC project.
%Canada, China, India, Australia, and other
%nations are also being approached for participation in the ILC.

 Europe's technological expertise and its scientific strength make it
 a valued potential partner.   Japan is approaching Europe both
 through
bilateral discussions with individual countries, in which  ILC may
 appear in a broader landscape embracing other advanced technology
 topics, and through direct engagement with CERN. 
It is our hope  that CERN will play a leading
role in the European participation in the ILC, along the lines
described in the 
conclusions of the 2013 Update of the European Strategy, and also in a
similar fashion to that  developed for the European participation
in the US  neutrino program.  

ILC is an energy-frontier project that
can be started today.   It will provide a new opportunity for European
physicists in the time frame of the HL-LHC and beyond, as Europe plans
and marshals its resources for the next major CERN project. In this 
way, the ILC will play a crucial role in encouraging a new
generation of researchers to enter particle physics and maintain the
continuous tradition and the scientific strength of our enterprise.

In summary, a large world-wide community of particle physicists is eager to join the effort to build the ILC and its detectors, and to pursue its unique 
physics program. The machine technology is mature and construction-ready. The
envisaged timeline of the project includes 4 years of preparation
phase and 9 years of construction. The ILC will deliver unique contributions in our effort to probe beyond the Standard Model to an ultimate understanding
of the fundamental laws of nature. The scientific case for the ILC
has become irresistible.

%\section{\label{sec:sum}Summary \& Conclusion} 

%The recent construction of the European XFEL
%based on SCRF technology gives high reliability to the project
%including its industrialisation process and cost estimation. The
%detector R\&D puts the detector concepts on solid footing, ready to
%move to finalising choices of technologies and parameters, and
%optimising integration. The mature technology of the ILC
%means that it provides a dependable future beyond
%the LHC and the high luminosity upgrades HL-LHC. High-level 
%administrative and governmental bodies in Japan are evaluating
%and advancing the proposal to host the ILC. Some potential partner
%countries are being approached by Japan, and others will follow. 

\newpage
%\vspace*{3.0cm}

%\bibliographystyle{utphysmod}
\bibliography{ILC-ESU-refs}

\vspace{-.3cm}

\onecolumngrid
\newpage
%\vspace{1cm}

\appendix
%\part*{Appendices}

\addcontentsline{toc}{part}{Appendix}

\section*{\label{Appendix1} \Large{Appendix A: ILC250 project costs}} 

\begin{figure*}[ht]
 \begin{center}
 \includegraphics[width=18cm, height=11cm]{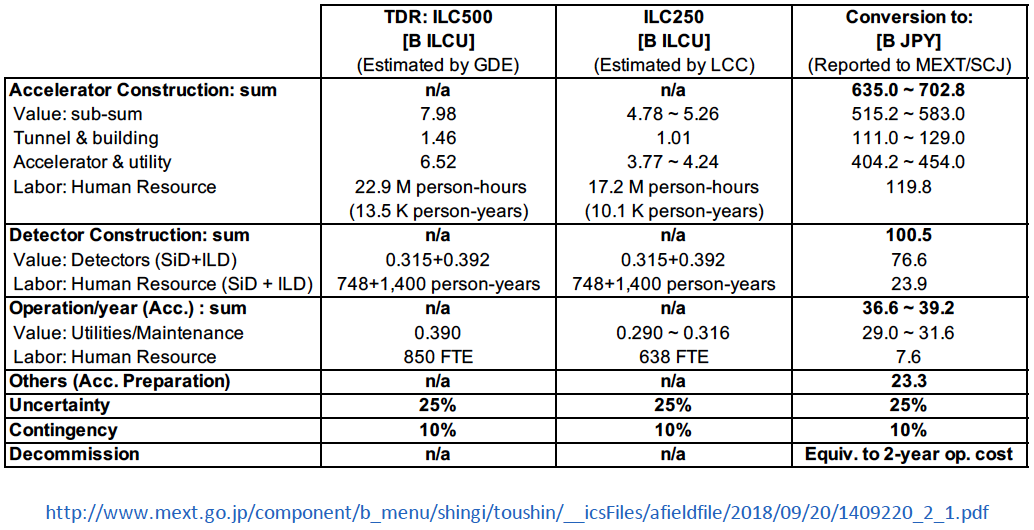}
\caption{Costs of the ILC250 project in ILCU as evaluated by the Linear Collider Collaboration (LCC), converted to JPY and re-evaluated by KEK, and summarised in the MEXT ILC Advisory Panel report, in July, 2018.
 \label{Cost}}
 \end{center}
 \end{figure*}

\textbf{The above summary is based on the information given in:} 
\begin{enumerate}
\item 
\textit{ILC-TDR} (note: it is a reference for the ILC500 in B ILCU): \url{http://www.linearcollider.org/ILC/Publications/Technical-Design-Report};
\item
\textit{The International Linear Collider Machine Staging Report 2017} (note: it is a reference for ILC250 cost in B ILCU): arXiv:1711.00568[hep-ex, \url{https://arxiv.org/abs/1711.00568};
\item
\textit{Summary of the ILC Advisory Panel's discussions to date after Revision} (conversion to JYen). Report by the International Linear Collider (ILC) Advisory Panel, MEXT, Japan on July 4, 2018: \url{http://www.mext.go.jp/component/b\_menu/shingi/toushin/\_\_icsFiles/afieldfile/2018/09/20/1409220\_2\_1.pdf}.
\end{enumerate}
 
The ILC currency unit (ILCU) is defined as 1 US Dollar (USD) in Jan., 2012. The cost conversion to Japanese Yen (JPY) has assumed that 1 Euro=115 JPY and 1 USD=100 JPY. The accelerator labor-estimate unit of ‘person-hours’ may be simply converted to ‘person-years’ by using a factor of 1,700 working-hours per year.

The total value cost for the 250 GeV accelerator construction was estimated to be in a range of 4.78-5.26 B ILCU and has been converted to 515.2-583.0 B JPY, by taking into account various effects of SCRF cost-reduction R\&D, smaller mass production because of ILC500 to ILC250, and time-dependent variations specially in tunnelling and building works. 

These numbers include the cost for civil engineering (tunnelling, building etc.) and the laboratory. Costs not included are land acquisition, living environment for visiting researchers, access roads, groundwater handling, energy service enterprise for power transmission, part of low power voltage supplies and physic-analysis computer centre. The cost premium to cover the project cost with 85\% instead of 50\% confidence level (loosely speaking, the 1 sigma  uncertainty of the cost estimate) has been estimated to be 25\% of the estimated cost.

%\begin{figure*}[h]
 %\epsfysize=9.0cm
 %\begin{center}
 %\includegraphics[width=7cm, height=8cm]{figures/Cost2.png}
%\caption{ \label{Cost2}}Cost(part A)of the project as presented by Sh. Michizono at LCWS 2018 as re-evaluated by the Japanese Ministry MEXT report.
 %\end{center}
 %\end{figure*}

%\begin{figure*}[hb]
 %\epsfysize=9.0cm
 %\begin{center}
% \includegraphics[width=11cm, height=7cm]{figures/AcceleratorCostDrivers.jpg}
%\caption{Breakdown of major cost drivers of the acclerator. %\label{AcceleratorCostDrivers}}
% \end{center}
% \end{figure*}
%\vspace*{5cm}
\newpage

\section*{\label{Appendix2} \Large{Appendix B: Definition of the Community}} 
\begin{figure*}[h]
 \begin{center}
 \includegraphics[width=\hsize,height=9cm]{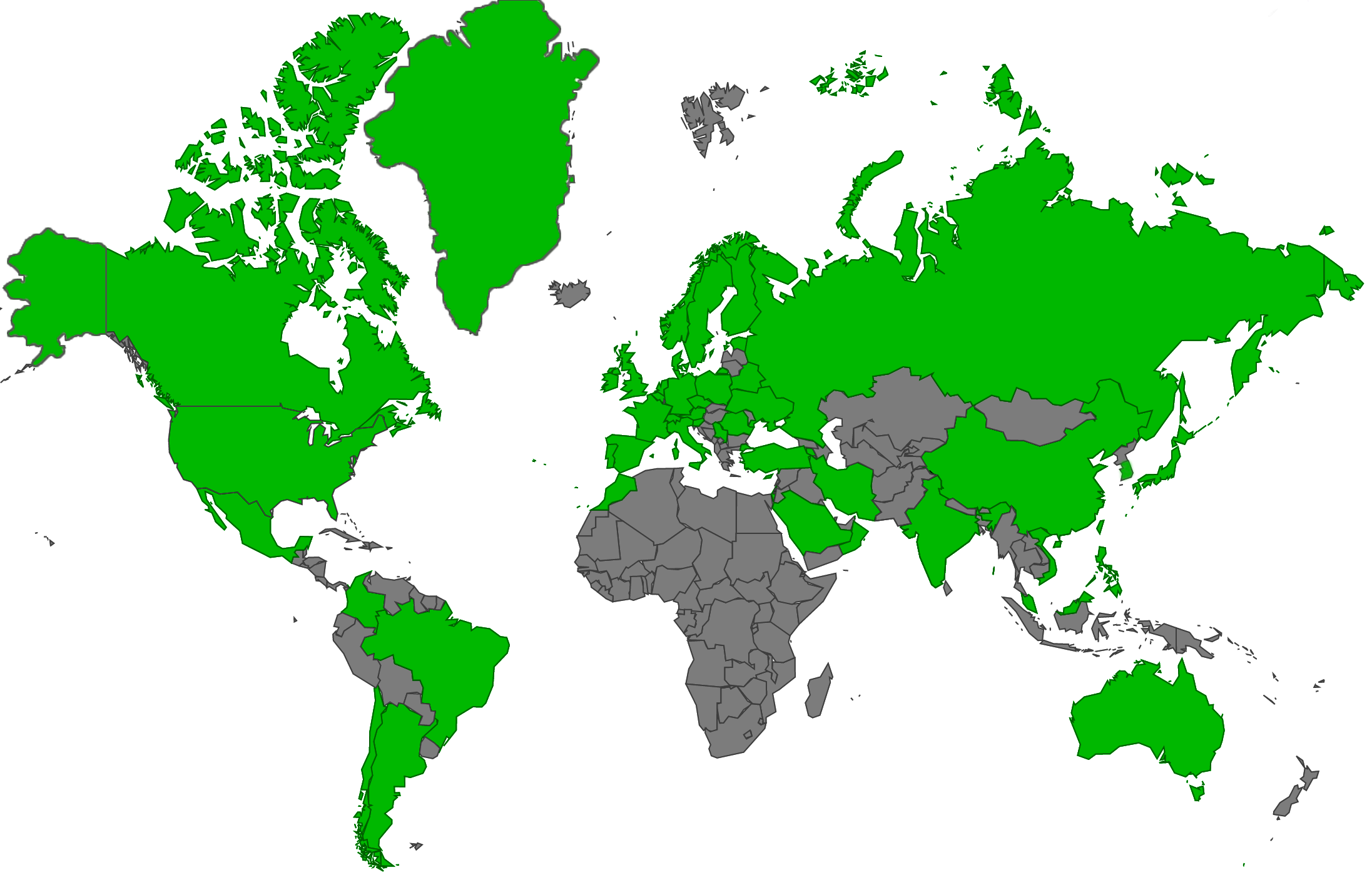}
\caption{World wide map distribution of signatories supporting the ILC Technical Design Report. \label{TDRsignatories}}
 \end{center}
 \end{figure*}
 
 \begin{figure*}[h]
 \begin{center}
 \includegraphics[width=\hsize,height=8cm]{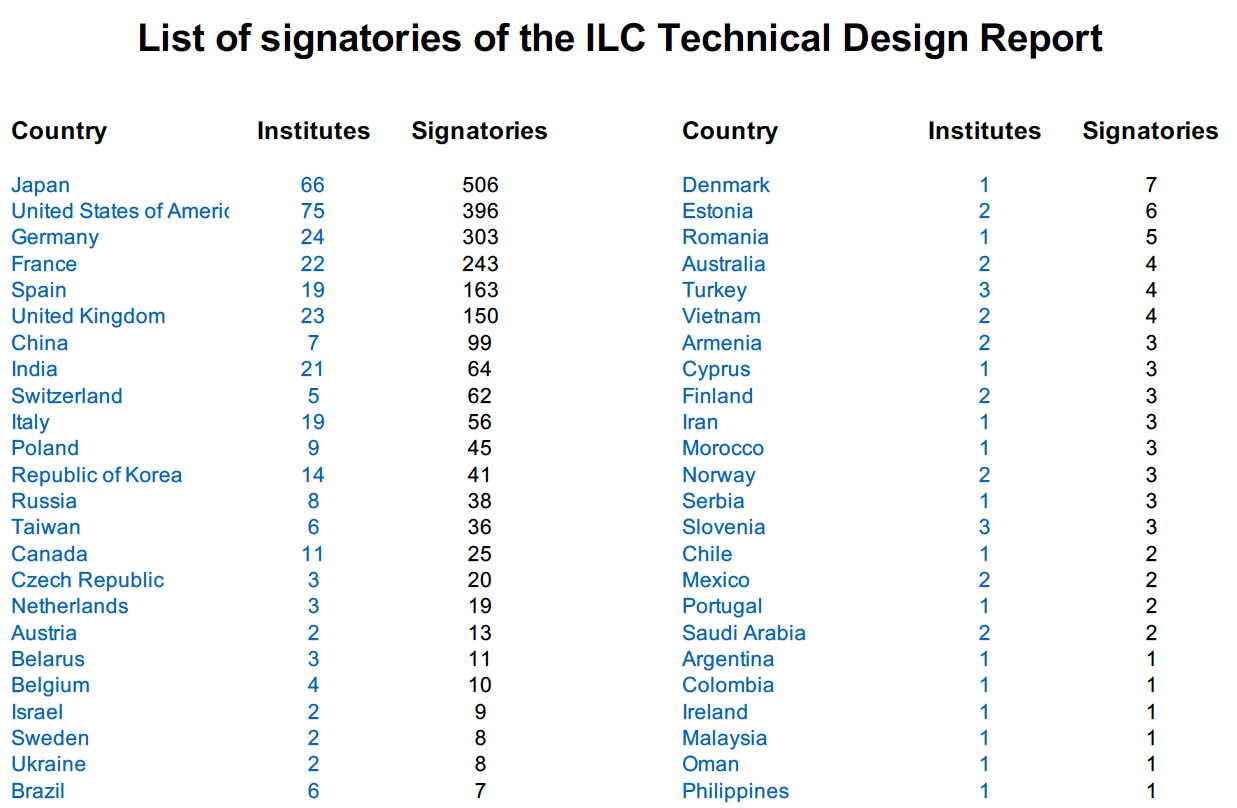}
\caption{Detailed list of signatories of the ILC Technical Design Report covering 2400 signatories, 48 countries and 392 Institutes/Universities. \label{TDRsignatories}}
 \end{center}
 \end{figure*}

\newpage

\section*{\label{Appendix3} \Large{Appendix C: List of supporting documents} }
%Description of supporting documents:
\begin{itemize}
\item
ILC TDR documents;
\item
ILC general overview, being specifically produced for the European Strategy Process;
\item
European ILC Preparation Plan (EIPP), produced under the E-JADE project;
\item
Linear collider Detectors R\&D Liasion Report;
\item
Green ILC project: reports and web page;
\item
Letter from the KEK’s ILC Planning Office.

\end{itemize}

\textbf{Supporting documents web page:} 

%https://linearcollider.web.cern.ch/content/ilc-european-strategy-document

\url{https://ilchome.web.cern.ch/content/ilc-european-strategy-document}

\section*{\label{Appendix4} \Large{Appendix D: Glossary} }
%Abbreviations and definitions used in the text:
\begin{itemize}
\item
\textbf{AAA:} The Japanese Advanced Accelerator Association promoting science and technology (\url{http://aaa-sentan.org/en/association/index.html}).
\item
\textbf{AIDA:} Advanced European Infrastructures for Detectors at Accelerators. AIDA was funded by the EU under FP7 (\url{https://aida-old.web.cern.ch/aida-old/index.html}).
\item
\textbf{AIDA-2020:} Advanced European Infrastructures for Detectors at Accelerators. The successor of AIDA; AIDA-2020 is funded by the EU under Horizon2020 (\url{http://aida2020.web.cern.ch/}).
\item
\textbf{CALICE Collaboration:} R\&D group of more than 280 physicists and engineers from around the world, working together to develop a high granularity calorimeter system optimised for the particle flow measurement of multi-jet final states at the ILC running, with centre-of-mass energy between 90 GeV and ~1 TeV (\url{https://twiki.cern.ch/twiki/bin/view/CALICE/WebHome}).
\item
\textbf{CARE:} Coordinated Accelerator Research in Europe. CARE was funded by the EU under the FP6 programme.
\item
\textbf{E-JADE:} The Europe-Japan Accelerator Development Exchange Programme. E-JADE is a Marie Sklodowska-Curie Research and Innovation Staff Exchange (RISE) action, funded by the EU under Horizon2020 (\url{https://www.e-jade.eu/}).
\item
\textbf{EUDET:} Detector R\&D towards the International Linear Collider. EUDET was funded by the EU under the FP6 programme (\url{https://www.eudet.org/}).
\item
\textbf{European XFEL:} The European X-Ray Free-Electron Laser Facility (European XFEL) at DESY (Hamburg, Germany) (\url{https://www.xfel.eu}).
\item
\textbf{EUROTeV:} European Design Study Towards a Global TeV Linear Collider. EUROTeV was funded by the EU under the FP6 programme (\url{https://www.eurotev.org/}).
\item
\textbf{ICFA:} International Committee for Future Accelerators (http://icfa.fnal.gov/).
\item
\textbf{ILC-HiGrade:} International Linear Collider and High Gradient Superconducting RF-Cavities. ILC-HiGrade was funded by the EU under the FP7 programme (\url{https://www.ilc-higrade.eu/}).
\item
\textbf{JAHEP:} Japanese Association of High Energy Physics.
\item
\textbf{Japanese National DIET:} The National Diet is Japan's bicameral legislature. It is composed of a lower house called the House of Representatives, and an upper house, called the House of Councillors.
\item
\textbf{LCLS-II:}  The hard X-ray free-electron laser at SLAC (Stanford, USA)(\url{https://portal.slac.stanford.edu/sites/lcls-public/lcls-ii/Pages/default.aspx}).
\item
\textbf{MEXT:} Ministry of Education, Culture, Sports, Science and Technology (\url{http://www.mext.go.jp/en/}).
\item
\textbf{SHINE:} Hard X-Ray free electron laser facility in Shanghai.

\end{itemize}

\end{document}